\documentclass[twocolumn,aps,showpacs]{revtex4}

\usepackage{amsmath}
\usepackage{amssymb}
\usepackage{graphicx}
\usepackage{dcolumn}

\begin{document}

\title{Effect of Landau quantization on linear magnetoresistance of periodically modulated 
two-dimensional electron gas.}

\author{O. E. Raichev}
\affiliation{Institute of Semiconductor Physics, National Academy of Sciences of Ukraine, 
Prospekt Nauki 41, 03028, Kyiv, Ukraine}
\date{\today}

\begin{abstract}
The linear response of two-dimensional electron gas in a perpendicular magnetic field in 
the presence of a spatially dependent classically smooth electrostatic potential is studied 
theoretically, by application of the Kubo formula for nonlocal conductivity tensor. In the 
classical transport regime, a general expression for the conductivity tensor through the 
correlation functions of the homogeneous electron gas is derived. The quantum transport regime, 
when Landau quantization is essential, is studied for the case of unidirectional periodic 
potential modulation. Apart from the Shubnikov-de Haas oscillations, the 
resistivity can demonstrate quantum oscillations with larger periods and smaller amplitudes, 
which survive when temperature increases. These oscillations exist when the modulation 
amplitude considerably exceeds the cyclotron energy so the Landau subbands, formed out of 
the Landau levels by the modulation potential, overlap in the energy domain. Both diagonal 
components of the resistivity tensor demonstrate oscillations related to modification of the 
density of states by the modulation. In addition, the resistivity component perpendicular to 
the modulation axis, which is caused by the scattering-assisted hopping transport, shows 
another kind of oscillations related to enhancement of the hopping probability when the 
guiding center of cyclotron orbit shifts by the doubled cyclotron radius. It is suggested 
that such high-temperature oscillations can be detected under conditions when the modulation 
period considerably exceeds the cyclotron radius.
 
\end{abstract}

\pacs{73.43.Qt, 73.63.Hs, 72.10.Bg}

\maketitle

\section{Introduction}

Magnetotransport in two-dimensional (2D) electron gas is strongly influenced by the presence 
of a spatially varying electrostatic potential energy $U({\bf r})$ that describes either 
large-scale inhomogeneity of the system or intentional modulation introduced by different 
methods. The magnetoresistance of periodically modulated systems [1-53] demonstrates 
commensurability effects, in particular, Weiss oscillations in unidirectionally modulated 
2D electron gas, which have been thoroughly studied both experimentally and theoretically. 
These oscillations have classical origin [2], and they appear because of periodic 
dependence of the drift velocity, averaged over the path of cyclotron rotation, on the ratio 
of cyclotron radius $R$ to modulation period $a$. Similar oscillations exist in the 
case of periodic magnetic modulation created by a spatially varying component of the magnetic 
field. With increasing magnetic field, Landau quantization becomes important and the 
resistance shows quantum oscillations as well. 

Early experiments [1,3] employed a weak periodic modulation, whose amplitude was smaller 
than the cyclotron energy in the region of fields where Landau quantization was important. 
In this case, the quantum effects are basically reduced to the ordinary Shubnikov-de Haas 
oscillations (SdHO). Further experiments [17,22,26,27,33,42,43] with larger modulation 
amplitudes, employing either the potential modulation or the magnetic one, have demonstrated 
that SdHO are considerably modified by the modulation. In particular, a periodic variation 
of the SdHO amplitudes with magnetic field was observed. This effect is explained by a periodic 
variation of the density of states near the Fermi level due to the influence of modulation 
on the energy spectrum of electrons. Specifically, in the periodic unidirectionally 
modulated systems the Landau levels are transformed into one-dimensional Landau subbands 
whose bandwidths, as well as the shape of the corresponding density of states, oscillate with 
the subband number. This quantum-mechanical picture also was used for explanation of 
the classical Weiss oscillations, starting from Refs. [3,5,7]. The classical analog of the 
Landau subband spectrum is the dependence of the average of $U({\bf r})$ over the path of 
cyclotron rotation on the guiding center coordinate [4,7]. 

In spite of extensive studies of periodically modulated 2D electron gas in the past years, the 
theory of quantum magnetoresistance in such systems is still incomplete. In the previous theoretical 
works, calculations of magnetoresistance were based on the Kubo formula for local conductivity. 
However, a recent study [54] shows that it is necessary to start with the Kubo formula for 
nonlocal conductivity in order to obtain the results which are valid in a wide range of the 
parameter $R/a$ and conform with the results obtained from the Boltzmann equation formalism in 
the classical region of magnetic field $B$. Consequently, the nonlocal Kubo approach should be 
used in the quantum region of $B$ as well, and this means that the problem of quantum 
magnetotransport in periodically modulated 2D electron gas needs to be revisited. Next, more 
work is required for the systems with large modulation amplitudes, as the existing theoretical 
studies [27,33,43] of magnetotransport is such systems are limited. In particular, the effect 
of transitions of electrons between the Landau levels has not been studied systematically. 
Meanwhile, such transitions are important in magnetotransport because they can lead to 
magnetoresistance oscillations of resonance nature, which, unlike the SdHO, are not 
related to the position of the Fermi level with respect to the Landau levels and, for this reason, 
are not strongly suppressed by the temperature $T$. Several kinds of such oscillations have been 
found in high-mobility 2D electron gas at moderately strong magnetic fields below 1 Tesla [55]. 
The resonance transitions between the Landau levels in spatially homogeneous 2D systems occur 
both under quasi-equilibrium transport conditions, due to inelastic scattering by phonons, and 
under non-equilibrium conditions implying either a large current that leads to a tilt of the 
Landau levels by the Hall electric field or microwave irradiation that leads to photon-assisted 
scattering [55]. In modulated systems, such transitions do not require either inelastic 
scattering or non-equilibrium conditions, because the Landau levels are tilted by the modulation 
potential itself (see Fig. 1). The underlying physics is explained below in more detail.

\begin{figure}[ht]
\includegraphics[width=9.cm]{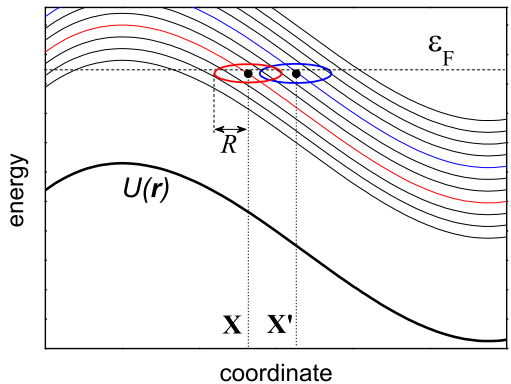}
\caption{(Color online) Landau quantization in the presence of a classically smooth potential 
energy $U({\bf r})$. For clarity, the potential is assumed to be one-dimensional, and only a part 
of the Landau levels is shown. An electron belonging to the Landau level $N$ and orbiting around 
the guiding center ${\bf X}$ sweeps during its cyclotron motion through several other Landau 
levels and can jump to the state with guiding center ${\bf X}'$ of another Landau level $N'$ 
as a result of elastic scattering near the Fermi level. Such transitions have a substantial 
influence on the resistance.}
\end{figure}

Consider the regime of classically strong magnetic fields, when the cyclotron frequency 
$\omega_c$ is much larger than the transport scattering rate $1/\tau_{tr}$. In high-mobility 
2D systems based on GaAs quantum wells, this regime is typical, as it starts already in the 
magnetic fields smaller than 0.1 Tesla. Then, the electronic motion in the 
presence of potential $U({\bf r})$ is subdivided into a fast cyclotron component and slower 
components including a drift caused by the electric field ${\bf E}({\bf r}) 
=-e^{-1} \partial U({\bf r})/\partial {\bf r}$, where $e$ is the electron charge, and a 
diffusion caused by scattering. Each mode of the fast degree of freedom corresponds to a 
different Landau level, while the slow degrees of freedom can be viewed as a drift of the 
guiding center in the crossed magnetic and electric fields and random jumps of this 
center when the electron changes its direction of motion in the scattering processes. The latter 
are determined mostly by the elastic impurity-assisted scattering if $T$ is sufficiently 
low. For weak and smooth potentials (see the conditions in Eq. (2) below), the Landau level number 
$N$ and the guiding center coordinate ${\bf X}$ can be considered as good quantum numbers, similar 
to the homogeneous case when the potential $U({\bf r})$ is absent. In the general case, the electron 
associated with a guiding center feels an effective potential averaged over the path of cyclotron 
rotation [56]. With increasing magnetic field, the limit of adiabatic motion is reached, which 
means that the relative change of the potential on the scale of cyclotron radius becomes small, so 
the electron feels the local potential $U({\bf X})$ and is characterized by a local drift velocity 
proportional to ${\bf E}({\bf X})$.  

Even within the quasiclassical picture of motion described above, the problem of magnetotransport 
appears to be essentially nontrivial if a variation of the potential energy $U({\bf r})$ on 
the scale of cyclotron orbit diameter considerably exceeds the Landau level separation $\hbar \omega_c$ 
(see Fig. 1). An electron in the Landau level $N$, orbiting around the guiding center ${\bf X}$, passes 
through the region where the states belonging to the other Landau levels exist at the same energy. 
Therefore, the electron can scatter, even elastically, into another Landau level, and the drift-diffusion 
motion of the guiding centers, which contributes to conductivity, is generally accompanied with transitions 
between Landau levels. This property causes a suppression of the SdHO and, more importantly, can lead 
to other kinds of magnetooscillation phenomena if transitions between the Landau levels have a resonance 
nature. In the case of periodic modulation, the existence of elastic transitions between the 
Landau levels implies that the doubled amplitude $2 u$ of the potential energy is larger than the 
Landau level separation. If $\hbar \omega_c$ is much smaller than the Fermi energy $\varepsilon_F$ 
(only this situation is considered below), the necessary condition is still achievable under the 
strong inequality $u \ll \varepsilon_F$ satisfied in all experiments on periodically modulated 2D systems.    

Below it is shown (see also Ref. [54]) that in the classical limit, when Landau quantization 
is neglected, application of nonlocal Kubo formalism allows one to express the conductivity 
tensor of a weakly modulated 2D system through the correlation functions of a homogeneous (unmodulated) 
system, which are calculated analytically in the case of isotropic elastic scattering. This result can 
be applied to any classically smooth potential $U({\bf r})$. In particular, for one-dimensional periodic 
potential one obtains an expression for magnetoresistance consistent with that derived from 
the Boltzmann equation in the theory of Weiss oscillations [2,20,21]. The magnetoresistance in 
two-dimensional periodic potential demonstrates similar commensurability oscillations.

The nonlocal Kubo approach applies to the quantum region of magnetic fields as well, though the 
conductivity is no longer reduced to the correlation functions of a homogeneous system. The 
quantum transport regime is studied in this paper for a particular case of unidirectional 
periodic modulation. The calculation shows that the resistivity retains weak quantum oscillations 
at elevated temperatures, when SdHO are completely suppressed. The resistivity components $\rho_{xx}$ 
and $\rho_{yy}$ (along and perpendicular to the modulation axis, respectively), in general, 
demonstrate oscillations of different origin. The resistivity along the modulation axis 
oscillates as a periodic function of the ratio of the Landau subband width to the cyclotron 
energy $\hbar \omega_c$, basically following slow oscillations of the density of states caused 
by the modulation. These weak oscillations of $\rho_{xx}$ correlate with the amplitude modulation of the 
SdHO discussed in the previous studies [26,27,33,42,43]. The resistivity perpendicular to the modulation 
axis shows oscillations with the same periodicity only in the region of low $B$, where $R > a/4$. 
They occur because of periodic resonance enhancement of the scattering between different Landau 
subbands when the maxima of the density of states in these subbands become aligned in energy. 
As the maxima are placed at the upper and lower edges of the Landau subbands, 
the subband width plays the role of the resonance energy. With increasing magnetic 
field and modulation period $a$, these oscillations disappear and another kind of oscillations 
emerges, whose periodicity is well-defined in the adiabatic limit $R \ll a$ and determined by 
the ratio $2 |e| {\overline E} R/\hbar \omega_c$, where ${\overline E}$ is approximately equal to 
the amplitude of electric field created by the modulation. Such oscillations are similar to 
those observed in nonlinear transport in homogeneous 2D systems [57-67], 
with the difference that ${\overline E}$ is replaced by the Hall electric field induced by the electric 
current. The resonance energy $2 |e| {\overline E} R$ defines a variation of electrostatic 
potential energy on the scale of cyclotron diameter $2R$ and originates from the property of enhanced 
backscattering in 2D systems: the scattering probability as a function of the momentum transferred 
in the transition has a maximum when this momentum is close to the doubled Fermi momentum $2 p_F$ 
so that the guiding center shifts by $2R$. The difference in the oscillating 
behavior of the components $\rho_{xx}$ and $\rho_{yy}$ described above is caused by two 
reasons. The first one is the difference between the hopping transport and the band transport, as 
the latter largely contributes into $\rho_{xx}$ and does not contribute into $\rho_{yy}$, and the 
second one is the anisotropy of the hopping transport.

The paper is organized as follows. Section II contains the details of calculation of the nonlocal 
conductivity tensor. In Sec. III, the classical limit is considered, and the general solution and 
its applications are discussed. In Sec. IV, the quantum contributions to the conductivity are 
calculated for the case of one-dimensional periodic modulation. Section V presents plots of the 
resistivity components versus the magnetic field, their detailed discussion, and concluding 
remarks.

\section{General formalism}

In the following, the Planck's constant $\hbar$ is set at unity. A parabolic spectrum of free electrons 
is assumed and the Zeeman splitting is neglected, so the electron states are doubly degenerate in spin. 
The Hamiltonian of noninteracting 2D electrons in a perpendicular magnetic field ${\bf B}=(0,0,B)$ has 
a standard form, $\hat{H}=\sum_j \hat{H}_{{\bf r}_j}$, where the sum is taken over all electrons, with 
a single-electron Hamiltonian  
\begin{equation}
\hat{H}_{{\bf r}}=\frac{1}{2m} \left( -i \frac{\partial}{\partial {\bf r}} - 
\frac{e}{c} {\bf A}_{\bf r} \right)^2 + U({\bf r}) + V({\bf r}). 
\end{equation} 
In this expression, ${\bf r}=(x,y)$ is the 2D coordinate, $m$ is the effective mass of electron, 
${\bf A}$ is the vector potential describing the magnetic field, and $V({\bf r})$ is a random 
potential energy due to impurities or other static inhomogeneities. It is assumed that $V({\bf r})$ 
varies on a scale much smaller than the cyclotron radius.  
The modulation potential energy $U({\bf r})$ is assumed to be weak, its amplitude $u$ is much 
smaller than the chemical potential (Fermi energy) $\varepsilon_F$, and classically smooth, which means 
that the spatial scale of $U({\bf r})$, estimated in the case of periodic modulation as the half-period 
$a/2= \pi/Q$, where $Q$ is the modulation wavenumber, is much larger than the magnetic length 
$\ell=\sqrt{c/|e|B}$. Furthermore, the drift velocity ${\bf v}_{D}({\bf r})= 
c [{\bf E}({\bf r}) \times {\bf B}] /B^2$ should be much 
smaller than the Fermi velocity $v_F=\sqrt{2 \varepsilon_F/m}$. This condition ensures that the 
drift-induced shift of the guiding center per one cyclotron rotation is much smaller than the 
cyclotron radius $R=v_F/\omega_c$, and can be rewritten in the form $\eta Q R \ll 1$, where $\eta 
= u/\varepsilon_F$ is the relative strength of the modulation. If $Q R < 1$, such a condition 
is always satisfied in view of $\eta \ll 1$. On the other hand, at $Q R > 1$ the drift of the 
guiding center is determined by the average drift velocity [2] that depends on the average potential 
acting on the electron during one cyclotron rotation [56,68]. As the amplitude of the average 
potential is reduced by a factor $\sqrt{Q R}$ compared to the amplitude of the actual potential 
[56], it is sufficient to assume a softer condition, namely $\eta \sqrt{Q R} \ll 1$. In summary, 
the necessary conditions applied throughout the paper are: 
\begin{equation}
\omega_c \ll \varepsilon_F,~  \ell \ll \pi/Q ,~ \eta \ll 1,~ \eta \sqrt{Q R} \ll 1. 
\end{equation} 

The steady-state nonlocal conductivity tensor is given by the Kubo-Greenwood formula, which is written 
below in the exact eigenstate representation:
\begin{eqnarray}
\sigma_{\alpha \beta}({\bf r},{\bf r}') = \frac{i}{S^2} \sum_{\delta \delta'} \frac{ \langle \delta'|
\hat{I}^{\alpha}_{{\bf r}}| \delta \rangle \langle \delta | \hat{I}^{\beta}_{{\bf r}'}| \delta' \rangle 
(f_{\varepsilon_{\delta}}-f_{\varepsilon_{\delta'}})}{(\varepsilon_{\delta}-\varepsilon_{\delta'}-i\lambda)
(\varepsilon_{\delta}-\varepsilon_{\delta'})},
\end{eqnarray}
where $\hat{{\bf I}}_{{\bf r}}=e\sum_j \{\hat{{\bf v}}_{{\bf r}_j},\delta({\bf r}_j-{\bf r})\}$ is the 
operator of current density expressed through the velocity operator $\hat{{\bf v}}_{{\bf r}}= 
[ -i \partial/\partial {\bf r} - (e/c){\bf A}_{{\bf r}} ]/m$, the curly brackets $\{,\}$ denote a symmetrized 
product, $\lambda \rightarrow +0$, $S$ is the normalization area, $\delta$ is the eigenstate index, and $f_{\varepsilon}=[e^{(\varepsilon-\varepsilon_F)/T} +1]^{-1}$ is the equilibrium Fermi distribution. 
It is convenient to transform Eq. (3) by applying the operator identity 
\begin{eqnarray}
\hat{{\bf v}}_{{\bf r}} = \ell^2 \hat{\epsilon} \frac{\partial {\cal U}_{\bf r}}{\partial {\bf r}} - 
\frac{i}{\omega_c} \hat{\epsilon} [\hat{{\bf v}}_{{\bf r}} ,\hat{H}_{\bf r} ],
\end{eqnarray}
where ${\cal U}_{\bf r}=U({\bf r}) + V({\bf r})$ is the total potential energy standing in the 
Hamiltonian (1) and $\hat{\epsilon}$ is the antisymmetric unit matrix in the Cartesian 2D 
coordinate space ($\epsilon_{xy}=-\epsilon_{yx}=1$, $\epsilon_{xx}=\epsilon_{yy}=0$). 
After substituting Eq. (4) into Eq. (3), the second term in Eq. (4) gives the non-dissipative 
classical Hall conductivity. The rest of the contributions come from the first term and are 
proportional to the products of the gradients of the total potential. The subject of interest 
is the dissipative part of the conductivity, which is written below through the Green's 
functions in coordinate representation:  
\begin{eqnarray}
\sigma^{d}_{\alpha \beta}({\bf r},{\bf r}') = 2 \pi e^2 \ell^4 \epsilon_{\alpha \gamma} 
\epsilon_{\beta \gamma'} \int d \varepsilon \left(- \frac{\partial f_{\varepsilon}}{\partial \varepsilon} \right)
\nonumber \\
\times \left< \frac{\partial {\cal U}_{{\bf r}}}{\partial r_{\gamma}} \frac{\partial {\cal U}_{{\bf r}'}}{\partial 
r'_{\gamma'}} {\cal A}_{\varepsilon}({\bf r},{\bf r}') {\cal A}_{\varepsilon}({\bf r}',{\bf r}) \right>. 
\end{eqnarray}
By convention, a summation over the repeating coordinate indices $\gamma$ and $\gamma'$ is implied. 
The angular brackets define the average over the random potential, and 
\begin{eqnarray}
{\cal A}_{\varepsilon}({\bf r},{\bf r}')=(2 \pi i)^{-1}[{\cal G}^A_{\varepsilon}({\bf r},{\bf r}') -
{\cal G}^R_{\varepsilon}({\bf r},{\bf r}') ]
\end{eqnarray}
is the spectral function in the coordinate representation, expressed through the nonaveraged 
Green's functions ${\cal G}^s$. The index $s$ denotes retarded (R) or advanced (A) Green's function. 
Since the case of degenerate electron gas is considered, the energy $\varepsilon$ stands in a narrow 
interval around the Fermi level, as defined by the energy derivative of the distribution function. 

\begin{figure}[ht]
\includegraphics[width=9.cm]{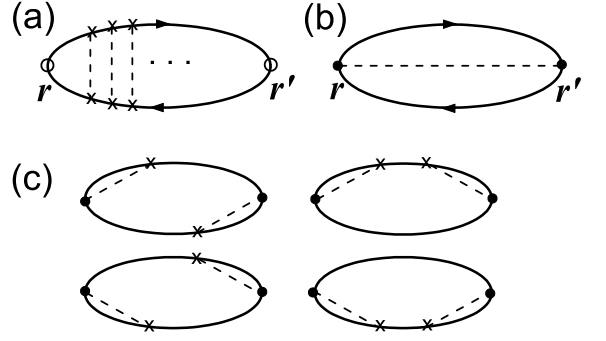}
\caption{Feynman diagrams representing different contributions to the nonlocal conductivity 
tensor. The crosses denote the random potential $V$, the filled circles denote its gradients in 
the points ${\bf r}$ and ${\bf r}'$, and the open circles in the diagram (a) denote gradients of 
the smooth potential $U$ (i.e., the electric fields $E$) in these points. The solid lines terminating 
at circles or crosses are the averaged Green's functions, while the dashed lines denote pair correlation 
functions of $V$. The diagram (a) actually implies a sum of infinite ladder series, beginning 
with the term without the dashed lines and then adding these lines one by one. The main 
contributions are given by the diagram (a) for $\sigma^{(1)}_{\alpha \beta}({\bf r},{\bf r}')$ 
and diagram (b) for $\sigma^{(2)}_{\alpha \beta}({\bf r},{\bf r}')$. The higher-order diagrams (c) 
lead to much smaller contributions under the condition $(\omega_c \tau_{tr})^2 \gg 1$.}
\end{figure}

In the average over the random potential, the mixed terms containing products of the gradients 
of $U({\bf r})$ and $V({\bf r})$ do not survive because $U({\bf r})$ and $V({\bf r})$ do not 
correlate with each other. The remaining terms can be evaluated within the accuracy up to the 
first power in the correlator $w(q)$ defined as a Fourier transform of the pair correlation 
function of the random potential. This leads to two distinct contributions [see Fig. 2 (a) and (b)], 
$\sigma^{d}_{\alpha \beta} \simeq \sigma^{(1)}_{\alpha \beta} + \sigma^{(2)}_{\alpha \beta}$, which 
are given by the following expressions:
\begin{eqnarray}
\sigma^{(1)}_{\alpha \beta}({\bf r},{\bf r}') = 2 \pi e^4 \ell^4 \epsilon_{\alpha \gamma} 
\epsilon_{\beta \gamma'} E_{\gamma}({\bf r}) E_{\gamma'}({\bf r}') 
\nonumber \\
\times \int d \varepsilon 
\left(- \frac{\partial f_{\varepsilon}}{\partial \varepsilon} \right) 
\left< {\cal A}_{\varepsilon}({\bf r},{\bf r}') {\cal A}_{\varepsilon}({\bf r}',{\bf r}) \right> 
\end{eqnarray}
and
\begin{eqnarray}
\sigma^{(2)}_{\alpha \beta}({\bf r},{\bf r}') = 2 \pi e^2 \ell^4
\epsilon_{\alpha \gamma} \epsilon_{\beta \gamma'} \int d \varepsilon 
\left(- \frac{\partial f_{\varepsilon}}{\partial \varepsilon} \right) 
\nonumber \\
\times \int \frac{d {\bf q}}{(2 \pi)^2} q_{\gamma} q_{\gamma'} w(q) e^{i {\bf q} \cdot ({\bf r}-{\bf r}')} 
A_{\varepsilon}({\bf r},{\bf r}') A_{\varepsilon}({\bf r}',{\bf r}), 
\end{eqnarray}
where $A_{\varepsilon}({\bf r},{\bf r}')= \left< {\cal A}_{\varepsilon}({\bf r},{\bf r}') \right>$ is 
the averaged spectral function. The first contribution, $\sigma^{(1)}$, describes the conductivity directly 
induced by the gradients of smooth potential, $\partial U({\bf r})/\partial {\bf r}=-e{\bf E}({\bf r})$. 
It differs from the Kubo-Greenwood expression for the dissipative part of the conductivity just by 
a formal substitution of the local drift velocity in place of the velocity operator. The second 
contribution, $\sigma^{(2)}$, is the leading term in the expansion of the conductivity in powers 
of the ratio of scattering rate to cyclotron frequency. Therefore, similar as in the case of 
unmodulated systems, $\sigma^{(2)}$ describes scattering-assisted 
hopping of electrons between the guiding centers of the cyclotron orbits.
Keeping the contributions (a) and (b) is sufficient in the case of classically strong magnetic 
fields. The diagram representation of the principal higher-order contributions is shown in 
Fig. 2 (c). In the self-consistent Born approximation (SCBA), the contributions shown in 
Fig. 2, complemented with the higher-order ones obtained from the diagrams (c) by adding 
possible noncrossing dashed lines, form a complete set for description of the conductivity.

One of the advantages of the approach based on the identity Eq. (4) is that the expression for 
conductivity tensor no longer contains matrix elements of the velocity operator, they are 
replaced by coordinate-dependent functions, the gradients of $U$ and $V$. Thus, there is no need to 
specify eigenstates and Green's functions on the early stage of calculation. Next, the diffusion-induced 
and drift-induced contributions are already separated. In particular, in the classical transport 
regime $\sigma^{(2)}$ describes the Drude conductivity while $\sigma^{(1)}$ is responsible for the 
commensurability oscillations. This makes a difference between the present technique and previous 
applications of the Kubo-Greenwood formalism to the problem. The most important difference, however, 
is a consideration of the nonlocal linear response instead of the local one. This is essential for 
evaluation of $\sigma^{(1)}$ as explained below. 
 
To find $\sigma^{(1)}$, one needs to calculate the pair correlation function entering Eq. (7) by 
considering the standard "particle-hole" ladder series, see Fig. 2 (a). In the case of arbitrary $w(q)$, 
the problem cannot be solved analytically even in the classical limit. This fact is consistent with the 
observation [21] that a solution of the Boltzmann equation with one-dimensional periodic potential 
cannot be presented in a closed analytical form for arbitrary $w(q)$. Therefore, the case of 
white noise random potential will be considered, when $w(q)$ is replaced by a constant 
so the scattering is isotropic. Introducing the correlation function $C^{ss'}_{\varepsilon}({\bf r},
{\bf r}')= w \langle {\cal G}^s_{\varepsilon}({\bf r},{\bf r}') {\cal G}^{s'}_{\varepsilon}({\bf r}',
{\bf r}) \rangle$ and applying a standard technique of the ladder series summation leads to the 
integral equation  
\begin{eqnarray}
C^{ss'}_{\varepsilon}({\bf r},{\bf r}')= K^{ss'}_{\varepsilon}({\bf r},{\bf r}')
+ \int d {\bf r}_1 K^{ss'}_{\varepsilon}({\bf r},{\bf r}_1) C^{ss'}_{\varepsilon}({\bf r}_1,{\bf r}'),
\end{eqnarray}
where $K^{ss'}_{\varepsilon}({\bf r},{\bf r}')= w G^s_{\varepsilon}({\bf r},{\bf r}') 
G^{s'}_{\varepsilon}({\bf r}', {\bf r})$ is the "bare" correlation function, which is expressed
through the averaged Green's functions $G^s_{\varepsilon}$ and corresponds to the diagram 
in Fig. 2 (a) without the dashed lines. Actually, only the terms with $s \neq s'$,
$C^{RA}_{\varepsilon}$ and $C^{AR}_{\varepsilon}$, are important in the pair correlation 
function of Eq. (7). It is convenient 
to rewrite Eq. (9) for the double Fourier transform $C^{ss'}_{\varepsilon}({\bf q},{\bf q}')=
\int d {\bf r} \int d {\bf r}' e^{-i {\bf q} \cdot {\bf r} + i {\bf q}' \cdot {\bf r}'} 
C^{ss'}_{\varepsilon}({\bf r},{\bf r}')$:
\begin{eqnarray}
C^{ss'}_{\varepsilon}({\bf q},{\bf q}')= K^{ss'}_{\varepsilon}({\bf q},{\bf q}')
\nonumber \\
+ \int \frac{d {\bf q}_1}{(2 \pi)^2} K^{ss'}_{\varepsilon}({\bf q},{\bf q}_1) 
C^{ss'}_{\varepsilon}({\bf q}_1,{\bf q}'),
\end{eqnarray}
where $K^{ss'}_{\varepsilon}({\bf q},{\bf q}')$ is the Fourier transform 
of $K^{ss'}_{\varepsilon}({\bf r},{\bf r}')$.

The correlator $C^{ss'}_{\varepsilon}$ essentially differs from $K^{ss'}_{\varepsilon}$. 
While $K^{ss'}_{\varepsilon}({\bf r}, {\bf r}')$ describes correlations on the scale of cyclotron 
diameter, $C^{ss'}_{\varepsilon}({\bf r},{\bf r}')$ has no definite correlation length and 
logarithmically depends on $|{\bf r}-{\bf r}'|$. This is a consequence of the diffusion-pole 
divergence of $C^{ss'}_{\varepsilon}({\bf q},{\bf q}')$ at $q \rightarrow 0$ and $q' \rightarrow 0$. 
Indeed, in the limit of small $q$ (in the classical transport regime $q \ll R^{-1}$ is sufficient) 
Eq. (10) can be reduced to a diffusion equation so that $C^{ss'}_{\varepsilon}({\bf 
r},{\bf r}')$ and $C^{ss'}_{\varepsilon}({\bf q},{\bf q}')$ are proportional to the Green's 
functions of the diffusion equation in the coordinate and momentum representation, respectively. 
The long-range behavior of correlations is a general property of 2D systems [69,70], which makes 
necessary the consideration of nonlocal conductivity. 

In contrast to $\sigma^{(1)}$, the contribution $\sigma^{(2)}$ can be treated locally, 
because it contains the exponential factor $e^{i {\bf q} \cdot ({\bf r}-{\bf r}')}$, where ${\bf q}$ 
has meaning of the momentum transferred in the scattering of electrons by the potential 
$V({\bf r})$. Since ${\bf q}$ is comparable to Fermi momentum (except for the case of 
scattering on very small angles), the correlation length $|{\bf r}-{\bf r}'|$ appears to be 
much smaller than both $R$ and $\pi/Q$, and it is sufficient to consider the local form
\begin{eqnarray}
\sigma^{(2)}_{\alpha \beta}({\bf r})= \int d \Delta {\bf r} \sigma^{(2)}_{\alpha \beta}({\bf r} + 
\Delta {\bf r}/2, {\bf r} - \Delta {\bf r}/2).
\end{eqnarray}    

A local approximation for $\sigma^{(1)}$ is possible when the modulation length is small enough 
so that relevant $q$ and $q'$ are much larger than $R^{-1}$, because in these conditions 
$K^{ss'}_{\varepsilon}({\bf q}, {\bf q}')$ becomes small and the integral term in Eq. (10) 
contains this smallness in a higher order. Therefore, Eq. (10) can be solved by iterations, and 
the zero iteration solution corresponds to a neglect of the integral term, when the exact 
correlator $C^{ss'}_{\varepsilon}$ is merely replaced by the bare correlator $K^{ss'}_{\varepsilon}$. 
In application to periodically modulated systems, this approximation transforms $\sigma^{(1)}$ into 
the band conductivity described in the previous theoretical works based on the local Kubo approach, 
starting from Refs. [3-5]; see the final part of Sec. IV for more details. 

In the case of a periodic potential $U({\bf r})$, the problem becomes macroscopically homogeneous 
and described by the conductivity tensor
\begin{eqnarray}
\sigma_{\alpha \beta}= \frac{1}{S} \int d {\bf r} \int d {\bf r}' \sigma_{\alpha \beta}({\bf r},
{\bf r}').
\end{eqnarray}
This conductivity can be also viewed as the average of the local conductivity, $\sigma_{\alpha 
\beta}({\bf r})$, over the elementary cell of the modulation lattice. It is important, 
however, that the calculation starts with the expression for nonlocal conductivity, and 
only when $\sigma_{\alpha \beta}({\bf r},{\bf r}')$ is found, which assumes calculation of 
the correlation function $C$ as described above, a transition to the form of Eq. (12) is 
carried out.

\section{Classical conductivity}

The contribution $\sigma^{(1)}$ is proportional to the squared gradient of $U({\bf r})$ and to
the Green's function correlators $C^{ss'}_{\varepsilon}$. Accounting for the presence 
of $U({\bf r})$ in the Green's functions leads to higher-order terms in expansion of $\sigma^{(1)}$ 
in powers of $U({\bf r})$ and $\nabla U({\bf r})$. In the quantum regime, when Landau quantization 
is important, this leads to the terms depending on $U({\bf r})/\omega_c$ and $\nabla U({\bf r}) 
R/\omega_c$ that cannot be neglected (see the next section). However, in the classical regime the 
expansion goes in the powers of small parameters $U({\bf r})/\varepsilon_F$ and $\nabla U({\bf r}) 
R/\varepsilon_F$. Therefore, to calculate $\sigma^{(1)}$ in the classical limit, it is sufficient 
to employ the averaged Green's function of a homogeneous system: 
\begin{eqnarray}
G^{s}_{\varepsilon} ({\bf r},{\bf r}') =\frac{e^{i \theta({\bf r},{\bf r}') }}{2 \pi \ell^2} 
\sum_{N=0}^{\infty} \frac{L_{N}(|\Delta {\bf r}|^2/2 \ell^2) e^{-|\Delta {\bf r}|^2/4 \ell^2}}{
\varepsilon - \varepsilon_N - \Sigma^{s}_{\varepsilon} },
\end{eqnarray}
where $\Delta {\bf r} ={\bf r}-{\bf r}'$, the sum is taken over Landau level numbers, $L_{N}$ 
are the Laguerre polynomials, $\varepsilon_N=\omega_c(N+1/2)$ is the Landau level spectrum, 
$\Sigma^{s}_{\varepsilon}$ is the self-energy, and $\theta({\bf r},{\bf r}')=(e/c) 
\int_{{\bf r}'}^{{\bf r}} d {\bf r}_{1} \cdot {\bf A}_{{\bf r}_{1}}$ is the gauge-sensitive 
phase. In the SCBA, the self-energy is determined from the equation $\Sigma^{s}_{\varepsilon}= 
(w/2 \pi \ell^2) \sum_{N} [\varepsilon - \varepsilon_N - \Sigma^{s}_{\varepsilon} ]^{-1}$, though
in the classical limit one can use $\Sigma^{A}_{\varepsilon}=-\Sigma^{R}_{\varepsilon}=i/2 \tau$,
where $\tau=1/mw$ is the scattering time. 

Because of the homogeneity of the problem, one has 
\begin{eqnarray}
K^{ss'}_{\varepsilon}({\bf q},{\bf q}')= K^{ss'}_{\varepsilon q} (2 \pi)^2 \delta({\bf q}-{\bf q}'), \nonumber \\
C^{ss'}_{\varepsilon}({\bf q},{\bf q}')= C^{ss'}_{\varepsilon q} (2 \pi)^2 \delta({\bf q}-{\bf q}'),
\end{eqnarray}
and Eq. (10) is solved as 
\begin{eqnarray}
C^{ss'}_{\varepsilon q}=\frac{K^{ss'}_{\varepsilon q}}{1-K^{ss'}_{\varepsilon q}}.
\end{eqnarray}
According to the definition of $K$ and Eq. (13),  
\begin{eqnarray}
K^{ss'}_{\varepsilon q}= \frac{w}{2 \pi \ell^2} \sum_{N,N'} \frac{(-1)^{N+N'}e^{-\beta} 
L_N^{N-N'}(\beta) L_{N'}^{N'-N}(\beta)}{(\varepsilon - \varepsilon_N -
\Sigma^s_{\varepsilon} )(\varepsilon - \varepsilon_{N'} - \Sigma^{s'}_{\varepsilon} ) }, 
\end{eqnarray}
where $\beta=q^2\ell^2/2$ and $L^{\alpha}_{N}(\beta)$ are the associated Laguerre polynomials. 
The classical limit corresponds to treatment of Landau level numbers as continuous variables 
and application of the asymptotic form of $L^{\alpha}_{N}(q^2\ell^2/2)$ at large $N$, keeping in mind 
that relevant $q$ is much smaller than the inverse quantum lengths since the case of classically 
smooth modulation is considered. Employing also $\Sigma^A_{\varepsilon}-\Sigma^R_{\varepsilon}=
i/\tau$, one obtains
\begin{eqnarray}
K^{RA}_{\varepsilon q}=K^{AR}_{\varepsilon q} \simeq \sum_{n=-\infty}^{\infty} 
\frac{J_n^2(q R_{\varepsilon})}{1+(n \omega_c \tau)^2}, \\
R_{\varepsilon} = \ell^2 p_{\varepsilon},~~~ p_{\varepsilon}=\sqrt{2 m \varepsilon}, \nonumber
\end{eqnarray}
where $J_n$ is the Bessel function and $R_{\varepsilon}$ is the cyclotron radius at the energy 
$\varepsilon$ (by definition, $R_{\varepsilon_F} = R$), expressed through the absolute value of 
electron momentum at this energy, $p_{\varepsilon}$.
Strictly speaking, Eq. (17) is valid when $|n|=|N-N'|$ is much smaller than $N$, though the 
inequality $N \gg 1$ and a rapid convergence of the series allow one to extend the range of 
$n$ to infinity. Moreover, if $(\omega_c \tau)^2 \gg 1$, it is sufficient to retain a single 
term with $n=0$, which leads to 
\begin{eqnarray}
C^{RA}_{\varepsilon q}=C^{AR}_{\varepsilon q} \simeq \frac{J_0^2(q R_{\varepsilon})}{1-
J_0^2(q R_{\varepsilon})}.
\end{eqnarray}
Noticing that only $C^{RA}_{\varepsilon q}$ and $C^{AR}_{\varepsilon q}$ are essential in the 
correlation function $\left< {\cal A}_{\varepsilon}({\bf r},{\bf r}') {\cal A}_{\varepsilon}(
{\bf r}',{\bf r}) \right>$ in Eq. (7), and taking into account that the electron gas is 
degenerate, $\varepsilon \simeq \varepsilon_F$, one obtains 
\begin{eqnarray}
\sigma^{(1)}_{\alpha \beta}({\bf r},{\bf r}') = \frac{e^2 \tau}{\pi m \omega_c^2}  \epsilon_{\alpha \gamma} 
\epsilon_{\beta \gamma'} \! \! \int \! \frac{d {\bf q}_1}{(2 \pi)^2} \! \int \! \frac{d {\bf q}_2}{(2 \pi)^2} 
\! \int \! \frac{d {\bf q}}{(2 \pi)^2} \nonumber \\
\times q_{1 \gamma} q_{2 \gamma'} U_{-{\bf q}_1} U_{{\bf q}_2} e^{i ({\bf q} -{\bf q}_1) \cdot {\bf r}}
e^{i ({\bf q}_2 -{\bf q}) \cdot {\bf r}'} \frac{J_0^2(q R)}{1-J_0^2(q R)},
\end{eqnarray}
where $U_{{\bf q}}$ is the Fourier transform of $U({\bf r})$. 

Next, application of the homogeneous Green's functions Eq. (13) to calculation of $\sigma^{(2)}_{\alpha 
\beta}({\bf r})$ in the classical limit gives a coordinate-independent isotropic Drude conductivity 
at $(\omega_c \tau)^2 \gg 1$:
\begin{eqnarray}
\sigma^{(2)}_{\alpha \beta}=\delta_{\alpha \beta} \frac{e^2}{2\pi m \omega_{c}^2 \tau} 
\int d \varepsilon \left(- \frac{\partial f_{\varepsilon}}{\partial \varepsilon} \right) p^2_{\varepsilon} 
= \delta_{\alpha \beta} \frac{e^2 n_s}{m \omega_{c}^2 \tau},
\end{eqnarray}
where $n_s$ is the electron density.
Calculating the contribution of the four diagrams in Fig. 2 (c) leads to an additional term  
$\sigma^{(3)}_{\alpha \beta}= - \sigma^{(2)}_{\alpha \beta}/[1+(\omega_{c} \tau)^2]$ that 
complements the conductivity to the full Drude form. A generalization of these results to 
the case of arbitrary $w(q)$ is straightforward and leads to a substitution of transport time 
$\tau_{tr}$, defined in a standard way, in place of $\tau$. The effect of the potential energy 
$U({\bf r})$ on $\sigma^{(2)}$ leads to a contribution of the order $(\omega_c \tau)^{-2} 
\sigma^{(1)}$ and can be neglected. Therefore, in the classical limit the contribution 
$\sigma^{(2)}$ does not depend on $U({\bf r})$. However, in the quantum transport regime 
$\sigma^{(2)}$ is essentially modified by the presence of $U({\bf r})$, as shown in the 
next section.

For any periodic modulation, application of Eq. (12) to Eq. (19) leads to the expression
\begin{eqnarray}
\sigma^{(1)}_{\alpha \beta}= \frac{e^2 \tau}{\pi m \omega_c^2} \epsilon_{\alpha \gamma} 
\epsilon_{\beta \gamma'} \int d {\bf q}
\frac{q_{\gamma} q_{\gamma'} \Omega_{{\bf q}} J_0^2(q R) }{1-J_0^2(q R)} ,
\end{eqnarray} 
with $\Omega_{{\bf q}}= \sum_{k_1,k_2} |U_{k_1,k_2}|^2 \delta({\bf q}- k_1 {\bf Q}_1 - k_2 {\bf Q}_2)$, 
where $k_1$ and $k_2$ are integers, ${\bf Q}_1$ and ${\bf Q}_2$ are the Bravais vectors 
of the reciprocal lattice, and $U_{k_1,k_2}$ are the Fourier coefficients of the periodic 
potential $U({\bf r})$. For harmonic unidirectional modulation, $U({\bf r})=u \cos(Qx)$,
the vectors are ${\bf Q}_1=(Q,0)$ and ${\bf Q}_2=(0,0)$, while nonzero elements are $U_{1, 0}=
U_{-1, 0}=u/2$. Thus, only the component $\sigma^{(1)}_{yy}$ survives, and it is identified with 
the Weiss oscillations term
\begin{eqnarray}
\sigma^{(1)}_{yy}= \frac{e^2 n_s \tau}{m} \left (\frac{\eta}{2} \right)^2 
\frac{(QR)^2 J_0^2(Q R)}{1-J_0^2(Q R)}.
\end{eqnarray} 
If $(\omega_c \tau)^2$ is not large, $J_0^2(Q R)$ should be replaced by $K^{RA}_{\varepsilon_F Q}$ 
from Eq. (17). The result Eq. (22) [see also the resistivity $\rho_{xx}$ of Eq. (24) derived from 
Eq. (22)] is in full accordance with the result of theories based on the Boltzmann equation [2,20,21]. 
Previous theories based on the Kubo formula for local conductivity miss the term $J_0^2(Q R)$ in the 
denominator. Within the formalism described in this paper, this would occur if the correlator 
$C$ were replaced by the bare correlator $K$ (see the discussion in the end of Sec. II). Such 
an approximation is sufficient at $Q R \gg 1$, when $J_0^2(Q R) \simeq (2/\pi Q R) \cos^2(Q R-\pi/4) 
\ll 1$, but becomes invalid at $QR < 1$, where $J_0^2(Q R) \simeq 1 -(QR)^2/2$.
  
For harmonic bidirectional rectangular modulation, $U({\bf r})=u_1 \cos(Q_1x)+u_2 \cos(Q_2y)$, one 
has ${\bf Q}_1=(Q_1,0)$, ${\bf Q}_2=(0,Q_2)$, and nonzero elements are $U_{1,0}=U_{-1,0}=u_1/2$, 
$U_{0,1}=U_{0,-1}=u_2/2$. This leads to a simple superposition of the Weiss oscillations of Eq. (22), 
with $\sigma^{(1)}_{yy} \propto u_1^2$ depending on $Q_1$ and $\sigma^{(1)}_{xx} \propto u_2^2$ 
depending on $Q_2$. A particular case is the symmetric square lattice with $u_1=u_2$ and 
$Q_1=Q_2$, for which $\sigma^{(1)}$ is isotropic. Similar results have been obtained in Ref. [9]. 
One should be careful, however, about the range of applicability of these results, because in the 
case of bidirectional modulation a drift of electrons along closed equipotential contours becomes 
important [29,31]. If the scattering that transfers electrons from these contours to other states 
is weak enough, the conductivity should be suppressed [25] and, moreover, a transition to stochastic 
motion of electrons is possible. The localization effects associated with the closed contours of motion 
are not described within the Born approximation applied in this paper, as well as within any 
perturbation-based approach. The problem of localization in electron transport under bidirectional 
modulation was discussed in more detail in Refs. [29,31,56]. 

Once the conductivity is known, the resistivity tensor $\rho_{\alpha \beta}$ is determined in a 
standard way by calculating the inverse of the conductivity tensor. If only the diagonal 
components of the dissipative conductivity exist (for example, in the case of unidirectional 
modulation along $x$, or bidirectional modulation along $x$ and $y$), the 
dissipative resistivity is also diagonal: $\rho_{xx}=\sigma^{d}_{yy}/(\sigma^2_H + \sigma^d_{xx} 
\sigma^d_{yy})$ and $\rho_{yy}=\sigma^{d}_{xx}/(\sigma^2_H +\sigma^d_{xx} \sigma^d_{yy})$, 
where $\sigma_H=e^2 n_s/m \omega_c$ is the Hall conductivity. For classically strong magnetic 
fields, the contribution $\sigma^{(2)}_{\alpha \alpha}$ is always much 
smaller than $\sigma_H$. Then, assuming that $\sigma^{(1)}_{\alpha \alpha}$ is also much 
smaller than $\sigma_H$, one has simply $\rho_{xx}=\sigma^{d}_{yy}/\sigma^2_H$ and $\rho_{yy}=\sigma^{d}_{xx}/\sigma^2_H$. Strictly speaking, this assumption is not always valid, 
because with increasing $B$ the contribution $\sigma^{(1)}_{\alpha \alpha}$ becomes larger than 
$\sigma^{(2)}_{\alpha \alpha}$ and may even exceed $\sigma_H$ under condition $\omega_c 
\tau \eta^2 > 1$, so the relation between the resistivity and conductivity becomes more 
complicated. Nevertheless, in the case of unidirectional modulation, when $\sigma^{(1)}_{xx}=0$, 
the product $\sigma^d_{xx} \sigma^d_{yy}$ is equal to $\sigma^{(2)}_{xx}(\sigma^{(1)}_{yy}+
\sigma^{(2)}_{yy})$ and is always much smaller than $\sigma^2_H$, in view of the third and 
the fourth strong inequalities of Eq. (2). Therefore, for unidirectional modulation along $x$ 
the resistivity components are
\begin{eqnarray}
\rho_{xx} \simeq (\sigma^{(1)}_{yy}+\sigma^{(2)}_{yy})/\sigma^2_H,~ \rho_{yy} \simeq 
\sigma^{(2)}_{xx}/\sigma^2_H,
\end{eqnarray}      
which is true as well in the quantum transport regime considered in the next section.
In the classical limit, according to Eqs. (20), $\sigma^{(2)}_{xx}/\sigma^2_H= 
\sigma^{(2)}_{yy}/\sigma^2_H= \rho_0$, where $\rho_0=m/e^2 n_s \tau$ is the zero-field 
resistivity. Then, according to Eq. (22), the classical resistivity is given by
\begin{eqnarray}
\rho_{xx} \simeq \rho_0 + \rho_0 \left (\frac{\eta}{2} \right)^2 
\frac{(v_F \tau Q)^2 J_0^2(Q R)}{1-J_0^2(Q R)},~  \rho_{yy} \simeq \rho_0,
\end{eqnarray}
The Weiss oscillations occur in the region $Q R > 1$, while in the region $Q R \ll 1$ the 
adiabatic limit is reached, where $\rho_{xx} \propto B^2$, in agreement with experiment [10].
  
The formalism developed above can be also extended to describe the classical magnetotransport 
in the cases of magnetic modulation and random modulation [54]. 

\section{Quantum conductivity}

The problem of classical conductivity studied in the previous section does not require consideration 
of the influence of potential energy $U({\bf r})$ on the spectrum and wave functions of electron 
system. When studying the quantum contribution, this influence should be specified in detail, which 
is done below for the case of classically smooth one-dimensional potential $U(x)$. 

\subsection{Green's function and density of states} 

After choosing the Landau gauge, ${\bf A}_{{\bf r}}=(0,Bx,0)$, and searching for the wave function in the absence 
of the scattering potential $V$ in the form $e^{i p y} \psi(x)$, where $p$ is the momentum along the $y$ 
axis, the eigenstate problem is reduced to a one-dimensional Schroedinger equation for $\psi(x)$. 
When the first two of the strong inequalities in Eq. (2) are satisfied, it is sufficient to apply 
quasiclassical methods [71] for solution of this problem. In particular, the energy spectrum can be 
found from the Bohr-Sommerfeld quantization rule by integrating the classical momentum between the 
turning points $x_1$ and $x_2$ for finite motion in the combined potential formed by a parabolic 
potential due to magnetic-field confinement and an additional potential $U(x)$:   
\begin{eqnarray}
\int_{x_1}^{x_2} \! dx \sqrt{2m [\varepsilon -U(x)] -\frac{(x-X)^2}{\ell^4}} = \pi \left(N+\frac{1}{2}\right), 
\end{eqnarray}
where $X=-\ell^2 p$ is the $x$-axis projection of the guiding center ${\bf X}=(X,Y)$.
In the case of weak $U(x)$ [the third of the strong inequalities in Eq. (2)], an
expansion of the integrand up to the first power of $U(x)$ is sufficient, and the 
spectrum is given by the following implicit equation:
\begin{eqnarray}
\varepsilon = \varepsilon_N + U_{\varepsilon X},~ \varepsilon_N=\omega_c (N+1/2), 
\end{eqnarray}
where
\begin{eqnarray}
U_{\varepsilon X} = \! \int_{-R_{\varepsilon}}^{R_{\varepsilon}} \! \!
dx \frac{U(X+x)}{\pi \sqrt{R_{\varepsilon}^2-x^2}} = \! \int_0^{\pi} \! \! \frac{d \varphi}{\pi}
U(X+R_{\varepsilon} \cos \varphi).
\end{eqnarray}
Since $X+R_{\varepsilon} \cos \varphi$ is the $x$-axis projection of the coordinate 
of electron rotating in a cyclotron orbit around the guiding center, the quantity 
$U_{\varepsilon X}$ is a classical expectation value of $U(x)$ or, equivalently, 
the average potential energy [4,7]. Finally, by noticing that $U_{\varepsilon X}$ slowly varies
with $\varepsilon$ on the scale of $\omega_c$ if the last strong inequality of Eq. (2) 
is satisfied, one may replace $U_{\varepsilon X}$ by $U_{NX} \equiv U_{\varepsilon_N X}$, 
which is equivalent to a substitution of the quantized cyclotron orbit radius, $R_{N}=\ell \sqrt{2N+1}$, 
in place of $R_{\varepsilon}$ in Eq. (27). For a particular case of periodic modulation with 
the period $a=2 \pi/Q$ and the symmetry $U(x)=U(-x)$, one has 
\begin{eqnarray}
U_{NX} = \sum_{l=-\infty}^{\infty} U_l J_0(l Q R_N) \cos(lQX), 
\end{eqnarray}
where $U_l$ are the Fourier coefficients of $U(x)$. For harmonic modulation, only the coefficients
$U_1=U_{-1}=u/2$ are nonzero. 

The electron energy spectrum 
\begin{eqnarray}
\varepsilon = \varepsilon_N + U_{NX}
\end{eqnarray}
is widely used for description of commensurability oscillations within the quantum linear 
response theory. Whereas in the present study $U_{NX}$ is identified with the average 
potential energy found from the 
Bohr-Sommerfeld quantization rule, most often $U_{NX}$ is explained as a first-order 
perturbation correction to the Landau quantization energy $\varepsilon_N$. Indeed, 
a calculation of the diagonal matrix elements of the potential with the Landau 
eigenstates $\psi(x)=\psi_{NX}(x)$ gives the result Eq. (28) at $N \gg 1$. Then Eq. (29) 
describes one-dimensional Landau subbands whose bandwidth, 
according to Eq. (28), oscillates as a function of the subband number. It is important 
to note that when the conditions of Eq. (2) are satisfied, Eq. (29) remains valid even 
if the amplitude $u$ of the potential $U(x)$ considerably exceeds the cyclotron energy 
so that several Landau subbands overlap in the energy domain. Some reasons why the first-order 
perturbation theory actually works in these conditions are described in the next paragraph.

By expanding the wave function in the full basis of Landau eigenstates, $\psi(x)= \sum_N 
b_{N}(X) \psi_{NX}(x)$, one obtains a set of linear equations
\begin{eqnarray}
(\varepsilon_N + U_{NX} - \varepsilon) b_N + \sum_{N'(N'\neq N)} U_{N N'}(X) b_{N'} =0, 
\end{eqnarray}
where $U_{N N'}=U_{N'N}$ are the nondiagonal matrix elements of $U(x)$. Equation (30) 
is exact. In the case of periodic potential, the quasiclassical approach gives  
$U_{NN'}(X) = \sum_{l} U_l J_{N-N'}(l Q R_{(N+N')/2}) \cos[lQX+\pi(N-N')/2]$, which 
can be obtained either from the asymptotic form of $\psi_{NX}(x)$ or from the general rule 
for calculation of quasiclassical matrix elements [71]. Since the spectrum is established in the 
form of Eq. (29), the contribution of the sum in Eq. (30) has to be negligibly small for all large $N$, 
which means that the diagonal approximation $\psi(x) \simeq \psi_{NX}(x)$ is valid. The mutual 
cancellation of the terms in the sum of Eq. (30) occurs because the quasiclassical matrix 
elements slowly change with $N+N'$ and rapidly change with $N-N'$. A numerical solution of 
the eigenstate problem Eq. (30), carried out for the case of harmonic potential, confirms 
that the spectrum (29) at large $N$ is a fairly good approximation whose accuracy rapidly 
improves with decrease of the parameter $\eta \sqrt{Q R_N}$. Note also that in the adiabatic 
limit $Q R_N \ll 1$ the eigenstate problem is reduced to an exactly solvable problem for 
electron in the crossed magnetic and electric fields, when the latter is constant and given by 
the gradient of $U(x)$ in the point $x=X$. The exact solution has the form of Landau eigenstate 
with a shifted guiding center, the small shift is proportional to drift velocity and can be 
safely neglected. Consistently, the nondiagonal matrix elements in Eq. (30) in the adiabatic 
limit are small as $(QR_N)^{|N-N'|}$ and can be neglected as well. The above consideration also 
shows that the wave functions of electrons can be taken as the ordinary Landau eigenstates 
$\psi_{NX}(x)$ within the accuracy of the approach.
 
\begin{figure}[ht]
\includegraphics[width=9.cm]{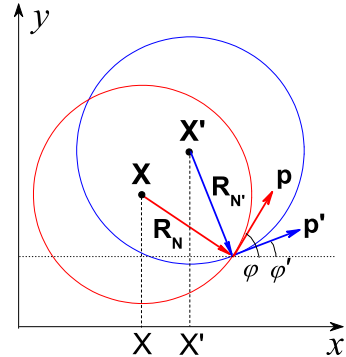}
\caption{(Color online) Scattering diagram for electrons moving in cyclotron orbits with radia $R_N$ 
and $R_{N'}$ around guiding centers ${\bf X}$ and ${\bf X}'$. The momentum ${\bf p}$ is uniquely 
related to the radius-vector ${\bf R}_N= \ell^2 \hat{\epsilon} {\bf p}$. The scattering rate of 
the electron belonging to the state $|N X \rangle$ into all other states $|N' X' \rangle$ is 
determined by a double integral over the angles of momenta ${\bf p}$ and ${\bf p}'$. Because 
of the presence of potential $U(x)$, the integrand depends on the difference $X'-X=R^{NN'}_{
\varphi \varphi'}$ between the projections of the guiding centers on the axis $x$.} 
\end{figure}

As the spectrum and eigenstates are specified, one can write the Green's function, averaged 
over the random potential $V$, in the following form: 
\begin{eqnarray}
G^{s}_{\varepsilon} ({\bf r},{\bf r}') = \sum_{N=0}^{\infty} 
\int \frac{d X}{2 \pi \ell^2} \frac{e^{-i X(y-y')/\ell^2} \psi_{N X}(x) \psi_{N X}(x') }{
\varepsilon - \varepsilon_N- U_{NX} - \Sigma^{s}_{\varepsilon NX}},
\end{eqnarray} 
where the self-energy, determined within the SCBA, is 
\begin{eqnarray}
\Sigma_{\varepsilon N X}^s = \sum_{N'=0}^{\infty} \int \frac{d {\bf q}}{(2 \pi)^2} 
\frac{w(q) \Phi_{NN'}(q^2 \ell^2/2)}{\varepsilon- \varepsilon_{N'}-U_{N' X'} 
- \Sigma^s_{\varepsilon N' X'} }, \\
X'=X + \ell^2 q_y, \nonumber
\end{eqnarray}
and $\Phi_{N N'}(\beta)= (N!/N^{\prime}!) \beta^{N^{\prime} -N} e^{-\beta} [L_{N}^{N^{\prime}-
N}(\beta)]^2$ is the squared matrix element of $e^{i{\bf q} \cdot {\bf r}}$ in the basis of the 
eigenstates $e^{-iXy/\ell^2}\psi_{NX}(x)$. In the quasiclassical case, the function $\Phi_{N N'}(\beta)$ 
rapidly oscillates with $\beta$ and exponentially rapidly decays at $\beta > 2(N+N'+1)$. It is 
sufficient to take into account only a smooth envelope of this function, which has the form 
$\Phi_{N N'}(\beta) \simeq [\pi \sqrt{\beta} \sqrt{2(N+N'+1)-\beta} ]^{-1}$ for $|N-N'| \ll N+N'$. 
Then Eq. (32) is rewritten as
\begin{eqnarray}
\Sigma_{\varepsilon N X}^s = \frac{1}{2 \pi \ell^2} \sum_{N'=0}^{\infty} \int_{0}^{2 \pi} 
\frac{d \varphi}{2 \pi} \int_{0}^{2 \pi} \frac{d \varphi'}{2 \pi} \nonumber \\ 
\times \frac{w(q^{NN'}_{\varphi \varphi'})}{\varepsilon-\varepsilon_{N'}-U_{N' 
X+R^{NN'}_{\varphi \varphi'}} -\Sigma^{s}_{\varepsilon N' X+R^{NN'}_{\varphi 
\varphi'}}  } , \\
R^{NN'}_{\varphi \varphi'}=R_N \sin \varphi-R_{N'} \sin \varphi', \nonumber
\end{eqnarray} 
where $q^{NN'}_{\varphi \varphi'}$ is approximately equal to 
$2 p_{\varepsilon} |\sin[(\varphi-\varphi')/2]|$. Equation (33) has a clear physical meaning 
(see Fig. 3). The self-energy describes the real and the imaginary corrections to electron energy 
because of electron scattering from the specified state $|N X \rangle$ into all other states. 
For electrons moving in cyclotron orbits, the scattering probability is formed by an integral over 
the angles of electron momenta in the initial and final states. The integrand depends on the scattering 
angle $\varphi-\varphi'$ through $w(q)$, because the scattering in general is sensitive to transferred 
momentum, and on the difference in guiding center projections, $R^{NN'}_{\varphi \varphi'}$, through 
the average potential. The maximum shift of the guiding center, $R_{N}+R_{N'}$, is realized for 
backscattering, when ${\bf p} \simeq -{\bf p}'$ and the cyclotron orbits touch each other in a 
single point.    

The equation for the self-energy generally requires a numerical solution. However, if the 
amplitude of quantum oscillations of $\Sigma_{\varepsilon N X}^s$ is small compared to $1/\tau$, 
it is sufficient to replace $\Sigma^{R,A}_{\varepsilon N' X+R^{NN'}_{\varphi \varphi'}}$ 
under the integral in Eq. (33) by $\mp i/2\tau$. When the amplitude of the 
average potential energy exceeds the cyclotron energy, this approximation is valid in a 
wider range of $B$ compared to unmodulated systems, because the modulation suppresses the
quantum oscillations of $\Sigma_{\varepsilon N X}^s$.
In the quasiclassical conditions, $U_{NX}$ slowly depends on Landau level number $N$ and the 
contribution to the sum in Eq. (33) comes mostly from a narrow interval of Landau levels 
near the energy $\varepsilon$. For this reason, $\Sigma_{\varepsilon N X}^s$ weakly depends 
on $N$ and can be denoted as $\Sigma_{\varepsilon X}^s$, assuming that $N \simeq \varepsilon/\omega_c-1/2$. 
Accordingly, one may replace all $N$ and $N'$ in Eq. (33) by $N=N'=\varepsilon/\omega_c-1/2$, except 
for the term $\varepsilon_{N'}$ which depends on $N'$ much stronger than $U_{N'X'}$. Similarly, 
$U_{NX}$ can be replaced by $U_{\varepsilon X}$, this equivalence is already discussed above. 
Finally, the model of isotropic scattering will be used, when $w(q)$ is a constant. An approximate 
quasiclassical expression for $\Sigma$ then takes the form
\begin{eqnarray}
\Sigma_{\varepsilon X}^A \simeq \frac{w}{2 \pi \ell^2} \sum_{N'=0}^{\infty} \int_{0}^{2 \pi} 
\frac{d \varphi}{2 \pi} \int_{0}^{2 \pi} \frac{d \varphi'}{2 \pi} \nonumber \\ 
\times \frac{1}{\varepsilon-\varepsilon_{N'}-U_{\varepsilon X+R_{\varphi \varphi'}} -i/2\tau } , \\
R_{\varphi \varphi'}=R_{\varepsilon} (\sin \varphi- \sin \varphi'), \nonumber
\end{eqnarray} 
and $\Sigma^R_{\varepsilon X}=\Sigma^{A*}_{\varepsilon X}$. In the case of harmonic modulation, 
$U(x)=u \cos(Qx)$, the calculation, based on the expansion of the integrand in the series of 
oscillating harmonics both in the energy and in the coordinate domains, leads to the following 
expression:  
\begin{eqnarray}
\Sigma^A_{\varepsilon X} = \frac{i}{2 \tau} \sum_{n=-\infty}^{\infty} \sigma_n e^{inQX},~
\sigma_n = \delta_{n,0}+2 \sum_{k=1}^{\infty} (-1)^{k}  \nonumber \\
\times d^{k} \exp \left( -i k \frac{2 \pi \varepsilon}{\omega_c} \right) i^n 
J_n(2 \pi k \tilde{u}_Q/\omega_c) J^2_{0}(n Q R_{\varepsilon}), \\
d=\exp (-\pi/\omega_c \tau),~ \tilde{u}_Q=u J_0(QR_{\varepsilon}), \nonumber
\end{eqnarray} 
where $d$ is the Dingle factor and $\tilde{u}_Q$ is the amplitude of the average potential $U_{\varepsilon X}$. 
The terms in the sum over $k$ describe quantum oscillations that are suppressed not only by the 
scattering but also by the smooth potential. Similar oscillations appear in the 
density of states. The average density of states $\rho_{\varepsilon}$ is given by the expression
\begin{eqnarray}
\rho_{\varepsilon} = \frac{2}{\pi S} {\rm Im} 
\int d {\bf r} G^{A}_{\varepsilon} ({\bf r},{\bf r})  = \frac{m}{\pi} \frac{1}{a} 
\int_0^{a} d X D_{\varepsilon}(X), \nonumber \\
D_{\varepsilon}(X) = \frac{\omega_c}{\pi} \sum_{N=0}^{\infty} 
{\rm Im} \frac{1}{\varepsilon - \varepsilon_N- U_{\varepsilon X} - \Sigma^{A}_{\varepsilon X}},
\end{eqnarray} 
where $D_{\varepsilon}(X)$ is the dimensionless (expressed in units $m/\pi$) local 
density of states, which is equal to unity in the classical limit. 
Being combined with Eqs. (27) and (34), the expression for $D_{\varepsilon}(X)$ is valid for arbitrary 
$U(x)$ and describes the density of states for electrons orbiting around the guiding centers with 
projection coordinate $X$. Under the approximation $\Sigma^A_{\varepsilon X} \simeq i/2 \tau$ and
for the case of harmonic modulation, one gets the result  
\begin{eqnarray}
D_{\varepsilon}(X) = \sum_{n=-\infty}^{\infty} \sum_{k=-\infty}^{\infty} (-1)^{k} 
d^{|k|} J_n(2 \pi k \tilde{u}_Q/\omega_c) \nonumber \\
\times \cos \left(k \frac{2 \pi \varepsilon}{\omega_c} -nQX- \frac{\pi n}{2} \right) 
\end{eqnarray} 
and
\begin{eqnarray}
\rho_{\varepsilon} = \frac{m}{\pi} \sum_{k=-\infty}^{\infty} (-1)^{k} d^{|k|} 
\cos \left(k \frac{2 \pi \varepsilon}{\omega_c} \right) J_0(2 \pi k \tilde{u}_Q/\omega_c).
\end{eqnarray}     
The average density of states in the form of Eq. (38) has been also obtained in Ref. [43]. This 
quantity describes equilibrium properties of the system but not the transport coefficients. As shown 
in the following subsections, the conductivity is determined by the local density of states 
$D_{\varepsilon}(X)$, which gives a more detailed description of the modulated 2D 
electron gas.

\subsection{Contribution $\sigma^{(2)}$} 

Consider the contribution $\sigma^{(2)}$ first. A substitution of the Green's function Eq. (31) 
into Eq. (8), with subsequent use of Eq. (11), leads to the following expression for the local 
conductivity:
\begin{eqnarray}
\sigma^{(2)}_{\alpha \alpha}(x) = \frac{e^2 m}{2 \pi} \int d \varepsilon 
\left(-\frac{\partial f_{\varepsilon}}{\partial \varepsilon} \right)  \nonumber \\
\times \int_{0}^{2 \pi} \frac{d \varphi}{2 \pi} 
\int_{0}^{2 \pi}  \frac{d \varphi'}{2 \pi} \frac{R^2_{\varepsilon} 
c_{\alpha}}{\tau} D_{\varepsilon}(X) D_{\varepsilon}(X'), \\
X=x-R_{\varepsilon} \sin \varphi,~~ X'=x-R_{\varepsilon} \sin \varphi', \nonumber 
\end{eqnarray} 
with $c_{x}=(\sin \varphi-\sin \varphi')^2$ and 
$c_{y}=(\cos \varphi -\cos \varphi')^2$, where $\varphi$ and $\varphi'$ are the angles of 
electron momenta. Equation (39) is valid for arbitrary (not necessarily periodic) $U(x)$ and 
describes the conductivity due to hopping transitions of electrons between the guiding centers 
with projection coordinates $X$ and $X'$. The hopping conductivity is proportional to the transition 
probability, expressed through the scattering rate $1/\tau$ and the product of the densities of states, 
and to the squared hopping distance $R^2_{\varepsilon} c_{\alpha}$ along the axis $\alpha$. Therefore, 
Eq. (39) can be viewed as a generalization of well-known Titeica's formula [72,73] to the case of 
electrons in a smooth potential, when the hopping is accompanied by transitions between the Landau 
levels. A combined effect of the potential and Landau quantization makes the hopping conductivity 
anisotropic. The theories of Refs. [33,43] lead to isotropic hopping conductivity because 
they do not account for higher-order quantum corrections (see below). An extension of 
Eq. (39) to arbitrary $w(q)$ is straightforward and implies a 
substitution of the angle-dependent scattering rate $\nu_{\varepsilon}(\varphi-\varphi') = 
mw(2 p_{\varepsilon} |\sin[(\varphi-\varphi')/2]|)$ in place of $1/\tau$.

The calculation of angular integrals in Eq. (39) is relatively simple under the approximation 
$\Sigma^A_{\varepsilon X} \simeq i/2 \tau$ and for harmonic modulation, $U(x)=u \cos(Qx)$. 
Then, after averaging over the period according to $\sigma^{(2)}_{\alpha \alpha}=a^{-1} \int_0^a 
\sigma^{(2)}_{\alpha \alpha}(x)$, one obtains the following expression:
\begin{eqnarray}
\sigma^{(2)}_{\alpha \alpha}= 
\sigma_d \sum_{n=-\infty}^{\infty} \sum_{k,k'=-\infty}^{\infty} (-1)^{k-k'}
d^{|k|+|k'|} \nonumber \\
\times {\cal T}_{k-k'} \exp \left[ i (k'-k) \frac{2 \pi \varepsilon_F}{\omega_c} \right] \nonumber \\ 
\times J_n(2 \pi k \tilde{u}_Q/\omega_c) J_n(2 \pi k' \tilde{u}_Q/\omega_c) {\cal B}_{\alpha}(nQR),
\end{eqnarray} 
where $\tilde{u}_Q$ is taken at $\varepsilon=\varepsilon_F$, $\sigma_d=e^2 n_s/m \omega_c^2 \tau$ is 
the classical Drude conductivity at $(\omega_c \tau)^2 \gg 1$, and ${\cal T}_k={\cal X}_k/\sinh 
{\cal X}_k$, with ${\cal X}_k=2 \pi^2 k^2 T/\omega_c$, is the thermal damping factor. 
The anisotropy is described by the functions
\begin{equation}
{\cal B}_{x}=J^2_0-J_0J_2 -2J^2_1,~{\cal B}_{y}=J^2_0+J_0J_2,
\end{equation} 
where the Bessel functions have the same argument as ${\cal B}_{\alpha}$. 
The classical conductivity contribution corresponds to $k=k'=0$. The principal harmonics 
of the SdHO come from the terms with $k=0$, $k'= \pm 1$ and $k= \pm 1$, $k'=0$. The terms 
with $k=k'\neq 0$ describe quantum corrections which are not suppressed by temperature and, 
therefore, are also important. Both the SdHO terms and the other quantum 
corrections show additional oscillations related to the presence of $U(x)$. These 
oscillations are described by the Bessel functions $J_n$ standing in Eq. (40) and by those 
entering ${\cal B}_{\alpha}$. When transport in high Landau levels is considered, it is 
often sufficient to keep only the principal SdHO harmonics together with $k=k'=\pm 1$ terms, 
which produces the following result:  
\begin{eqnarray}
\sigma^{(2)}_{\alpha \alpha} \simeq \sigma_d \left[ 1 - 4 d {\cal T}_1 {\cal J}_0
\cos \left( \frac{2 \pi \varepsilon_F}{\omega_c} \right) \right] + 
\delta \sigma^{(2)}_{\alpha \alpha}, \\
\delta \sigma^{(2)}_{\alpha \alpha} = \sigma_d 2 d^2 \sum_{n=-\infty}^{\infty} 
{\cal J}_n^2 {\cal B}_{\alpha}(nQR).   \nonumber
\end{eqnarray} 
Here and below, for the sake of brevity,
\begin{eqnarray}
{\cal J}_n \equiv J_n \left(\frac{2 \pi \tilde{u}_Q}{\omega_c} \right).
\end{eqnarray}
The SdHO term in Eq. (42) is proportional to $d$. It is isotropic, and its oscillations follow those 
of the density of states given by Eq. (38). The second-order quantum correction $\delta \sigma^{(2)}$, 
proportional to $d^2$, is anisotropic and describes transitions between the Landau levels. 

The case of very weak modulation, when $2 u \ll \omega_c$, corresponds to the situation when hopping 
transport is not affected by the presence of $U(x)$. Then, $\sigma^{(2)}$ is reduced to the conductivity 
of the homogeneous 2D electron gas, demonstrating the ordinary SdHO on the background of positive 
magnetoconductance. Formally, in this limit only a term with $n=0$ survives in the sum in Eq. (42), 
which leads to $\delta \sigma^{(2)}_{\alpha \alpha}/\sigma_d \simeq 2d^2$. The case of $2 u > \omega_c$ 
is far more interesting. Experimentally, it is found that the SdHO are considerably modified in this 
regime, showing the amplitude modulation with node points where the phase of SdHO is inverted 
[17,22,26,27,33,43]. This behavior is consistent with Eq. (42) as well as with the results of 
previous studies [26,27,33,43] based on simpler theoretical models. The quantum contribution $\delta 
\sigma^{(2)}$ has not been described in the previous theories. This contribution, however, is 
important, because it contains the oscillations that survive when temperature increases and 
SdHO disappear, see Sec. V.
 
In the adiabatic limit, $QR \ll 1$, it is more convenient to represent the term
$\delta \sigma^{(2)}_{\alpha \alpha}$ as an average of the local conductivity 
$\delta \sigma^{(2)}_{\alpha \alpha}(x)$ over the modulation period. This local 
conductivity is given by the following expression:
\begin{eqnarray}
\delta \sigma^{(2)}_{\alpha \alpha}(x)= \sigma_d 2 d^2 {\cal B}_{\alpha} 
\left( \frac{2 \pi e E(x) R}{\omega_c} \right),
\end{eqnarray} 
which is valid for arbitrary modulation and can be derived from Eq. (39) by expanding the densities 
of states in powers of $R_{\varepsilon}$. The high-temperature conductivity oscillations in this limit are 
described entirely by the oscillating properties of the functions ${\cal B}_{\alpha}$ defined by Eq. (41). 
The oscillations of ${\cal B}_{x}$ are much stronger than the oscillations of ${\cal B}_{y}$. This 
anisotropy exists because the transport along the modulation axis is much more often accompanied with
the hopping transitions between Landau levels than the transport perpendicular to this axis. Equation 
(44) is a particular case of a more general expression describing the $\propto d^2$ quantum 
correction to the local conductivity $\sigma^{(2)}$ for arbitrary potential $U({\bf r})$ 
and for arbitrary $w(q)$:
\begin{eqnarray}
\delta \sigma^{(2)}_{\alpha \beta}({\bf r}) = \frac{e^2 n_{s}}{m \omega_c^2} 2 d^2
\int_0^{\pi} \frac{d \theta}{\pi} (1  -  \cos \theta) \nu_{\varepsilon_F}(\theta) \nonumber \\
\times \left[ \delta_{\alpha \beta} J_0 \left(\frac{2 \pi e E({\bf r}) R}{\omega_c} 
2 \sin\frac{\theta}{2} \right) \right. \nonumber \\ 
\left. +J_2 \left(\frac{2 \pi e E({\bf r}) R}{\omega_c} 2 \sin\frac{\theta}{2} \right) 
\left(\delta_{\alpha \beta}-2 \frac{E_{\alpha}({\bf r}) E_{\beta}({\bf r}) }{E^2({\bf r})} 
\right) \right].
\end{eqnarray}
Thus, in the adiabatic limit the quantum contribution $\delta \sigma^{(2)}$ depends on the local 
potential through the gradient of this potential. The physics behind Eqs. (44) and (45) is 
described in more detail in Sec. V.  

\subsection{Contribution $\sigma^{(1)}$}

The nonlocal contribution $\sigma^{(1)}$ has only one component, 
$\sigma^{(1)}_{yy}({\bf r},{\bf r}')= \delta(y-y')\sigma^{(1)}_{yy}(x,x')$,
in the case of one-dimensional potential. The homogeneity of the system along the $y$ axis
implies that the correlator $C$ is representable in the form $C^{ss'}_{\varepsilon}({\bf q},
{\bf q}')=2 \pi \delta(q_y-q_y') C^{ss'}_{\varepsilon,q_y}(q_x,q'_x)$. The same representation 
is valid for the bare correlator $K$. Next, since only the correlators with $q_y=0$ 
enter $\sigma^{(1)}_{yy}(x,x')$, one needs to find $C^{ss'}_{\varepsilon}(q,q') \equiv 
C^{ss'}_{\varepsilon,0}(q_x,q'_x)$. Here and below $q_x$ is denoted as $q$ for brevity. 
Equation (10) is now rewritten as 
\begin{eqnarray}
C^{ss'}_{\varepsilon}(q,q')= K^{ss'}_{\varepsilon}(q,q') + \int \frac{d q_1}{2 \pi} 
K^{ss'}_{\varepsilon}(q,q_1) C^{ss'}_{\varepsilon}(q_1,q'),
\end{eqnarray}
where $K^{ss'}_{\varepsilon}(q,q')$ is obtained from $K^{ss'}_{\varepsilon}({\bf q},
{\bf q}')$ in the same way as $C^{ss'}_{\varepsilon}(q,q')$ is obtained from 
$C^{ss'}_{\varepsilon}({\bf q},{\bf q}')$. Only the terms with $s \neq s'$ (RA and AR) 
are important. Applying the Green's function of Eq. (31), one gets
\begin{eqnarray}
K^{RA}_{\varepsilon}(q,q')= w \sum_{N,N'}  \int \frac{d X}{2 \pi \ell^2} 
\int d x \int d x' e^{-iqx+iq'x'} \nonumber \\
\times \frac{\psi_{N X}(x) \psi_{N X}(x')}{\varepsilon - \varepsilon_N- U_{NX} - 
\Sigma^{R}_{\varepsilon NX}}  
\frac{\psi_{N' X}(x) \psi_{N' X}(x')}{\varepsilon - \varepsilon_{N'}- U_{N'X} - 
\Sigma^{A}_{\varepsilon N'X}}.
\end{eqnarray}
Similar to the homogeneous case considered in the previous section, only the terms 
with $N=N'$ are to be taken into account in the sum at $(\omega_c \tau)^2 \gg 1$, 
and $K^{RA}_{\varepsilon}=K^{AR}_{\varepsilon} \equiv K_{\varepsilon}$. By using 
asymptotic form of $\psi_{N X}(x)$ at $N \gg 1$ and taking into account that $q$ 
and $q'$ are small compared to the Fermi momentum, Eq. (47) is reduced to
\begin{eqnarray}
K_{\varepsilon}(q,q') \simeq  J_0(qR_{\varepsilon}) J_0(q'R_{\varepsilon}) 
\int dX e^{-i(q-q')X} \mu(X), 
\end{eqnarray}
where 
\begin{eqnarray}
\mu(X)=D_{\varepsilon}(X) \left[ \int_{0}^{2 \pi} \frac{d \varphi}{2 \pi} \int_{0}^{2 \pi} 
\frac{d \varphi'}{2 \pi} D_{\varepsilon}(X+R_{\varphi \varphi'}) \right]^{-1}.
\end{eqnarray}
To obtain Eqs. (48) and (49), the identity $2 {\rm Im} \Sigma^{A}_{\varepsilon X}= 
mw \int_{0}^{2 \pi} \frac{d \varphi}{2 \pi} \int_{0}^{2 \pi} \frac{d \varphi'}{2 \pi} 
D_{\varepsilon}(X+R_{\varphi \varphi'})$, based on a comparison of Eqs. (34) and (36), 
was applied. In the case of periodic modulation, $\mu(X)$ can be expanded in the 
Fourier series with coefficients $\mu_n$. As a result,
\begin{eqnarray}
K_{\varepsilon}(q,q')=2 \pi J_0(qR_{\varepsilon}) J_0(q'R_{\varepsilon}) \sum_n \delta(q-q'-nQ) \mu_n.  
\end{eqnarray}
Since $U(x)$ is real, $\mu^*_n=\mu_{-n}$ and $K^*_{\varepsilon}(q,q')=K_{\varepsilon}(q',q)$.
Below, the symmetry $U(x)=U(-x)$ is assumed, when the Fourier coefficients $U_n=U_{-n}$ are real, and 
so are $\mu_n$ and $K_{\varepsilon}(q,q')$. The function $\mu(X)$ becomes equal to unity, resulting 
in $\mu_n= \delta_{n,0}$, either in the absence of potential, when $D_{\varepsilon}$ is independent 
of coordinate, or in the classical case, when $D_{\varepsilon}=1$. This leads to the form 
$K_{\varepsilon}(q,q')=2 \pi \delta(q-q') J^2_0(qR_{\varepsilon})$ and to a simple
solution for $C_{\varepsilon}(q,q')$ exploited in the previous section. A combined 
effect of the potential and Landau quantization causes a significant dependence 
of $\mu(X)$ on $X$. 

The conductivity $\sigma^{(1)}_{yy}$ of a periodically modulated system involves only the 
terms with $q=nQ$ and $q'=n'Q$, where $n$ and $n'$ are integers. After introducing dimensionless 
coefficients $C_{n,n'}= L_x^{-1} C_{\varepsilon}(nQ,n'Q)$, where $L_x$ is the normalization length, 
and applying Eq. (12), this contribution is represented as
\begin{eqnarray}
\sigma^{(1)}_{yy}=\frac{e^2 \tau}{\pi m \omega_c^2} \sum_{n,n'} n n' Q^2 
U_{n} U_{n'} \int d \varepsilon \left(-\frac{\partial f_{\varepsilon}}{\partial 
\varepsilon} \right) C_{n,n'},
\end{eqnarray}  
where $C_{n,n'}$ is a solution of a set of linear equations
\begin{eqnarray}
C_{n,n'}= K_{n,n'} + \sum_{l} K_{n,l}C_{l,n'}.
\end{eqnarray}  
In this equation, $K_{n,n'}=J_0(nQR_{\varepsilon}) J_0(n'QR_{\varepsilon}) \mu_{n-n'}$ is 
a real symmetric matrix posessing also a symmetry $K_{n,n'}=K_{-n,-n'}$.
Since $\mu_{n-n'} = \delta_{n,n'} + \delta \mu_{n-n'}$, where $\delta \mu_{n-n'}$ denotes the 
quantum contribution, one has $K_{n,n'}=\delta_{n,n'} J_0^2(nQR_{\varepsilon})+ \delta K_{n,n'}$, 
where $\delta K_{n,n'}=J_0(n QR_{\varepsilon}) J_0(n'QR_{\varepsilon}) \delta \mu_{n-n'}$.

In the case of harmonic modulation, $U(x)=u \cos(Qx)$, it is convenient to introduce a 
function $F_n(\varepsilon)=(C_{n,1}-C_{n,-1}-C_{-n,1}+C_{-n,-1})/2$, which can be considered 
only for $n \geq 1$ in view of the symmetry $F_n(\varepsilon)=-F_{-n}(\varepsilon)$. 
For this function, Eq. (52) is rewritten as
\begin{eqnarray}
[1-J_0^2(nQR_{\varepsilon})] F_{n} -\sum_{l=1}^{\infty} M_{nl} F_{l} = \delta_{n,1} J_0^2(QR_{\varepsilon}) 
+ M_{n1},
\end{eqnarray}  
where $M_{nl}=\delta K_{n,l}-\delta K_{n,-l}$. Equation (51) is then rewritten as 
\begin{eqnarray}
\sigma^{(1)}_{yy}=\frac{e^2 n_s \tau}{m} \left (\frac{\eta}{2} \right)^2 (QR)^2 
\int d \varepsilon \left(-\frac{\partial f_{\varepsilon}}{\partial \varepsilon} \right) F_1(\varepsilon),
\end{eqnarray}    
and describes both the classical and the quantum contributions to the conductivity. Generally, 
Eq. (53) requires a numerical solution. However, assuming that the quantum contributions 
are small, one can solve Eq. (53) analytically by iterations. With the accuracy up to the second-order 
quantum terms, the solution is
\begin{eqnarray}
F_1(\varepsilon) \simeq \frac{J_0^2(QR_{\varepsilon})}{1- J_0^2(QR_{\varepsilon})} \left[1+ 
\frac{\delta \mu_{0}-\delta \mu_{2} }{1-J_0^2(QR_{\varepsilon})} \right. \nonumber \\
+ \left. \frac{1}{1-J_0^2(QR_{\varepsilon})} \sum_{n=1}^{\infty} 
\frac{(\delta \mu_{n-1}-\delta \mu_{n+1})^2 J_0^2(nQR_{\varepsilon})}{1-J_0^2(nQR_{\varepsilon})}
\right], 
\end{eqnarray}    
where the first term leads to the classical contribution. Within the required accuracy,
\begin{eqnarray}
\delta \mu_{l}= - 2 d [1-J_0^2(lQR_{\varepsilon}) ] 
{\cal J}_{l} 
\cos \left(\frac{2 \pi \varepsilon}{\omega_c} -\frac{\pi l}{2} \right) \nonumber \\
- 2 d^2 \cos \frac{\pi l}{2} \sum_{n=-\infty}^{\infty} {\cal J}_{n} {\cal J}_{n+l} \nonumber \\
\times J_0^2((n+l)QR_{\varepsilon}) [1-J_0^2(nQR_{\varepsilon})].
\end{eqnarray} 
Similar as in the previous subsection, only the principal SdHO harmonics should be taken into account 
so that a rapidly oscillating function of energy is retained only in the first term of this expression. 
Combining Eqs. (54), (55), and (56), one obtains 
\begin{eqnarray}
\sigma^{(1)}_{yy} \simeq
\sigma^{(1c)}_{yy} \left[1 - 2d {\cal T}_1 {\cal J}_{2} \cos \frac{2 \pi \varepsilon_F}{\omega_c} \right. \nonumber \\
\left. \times \frac{1-J_0^2(2QR)}{1-J_0^2(QR)}  \right] + \delta \sigma^{(1)}_{yy},
\end{eqnarray}    
where $\sigma^{(1c)}_{yy}$ denotes the classical conductivity given by Eq. (22). The second term in the square 
brackets of Eq. (57) describes the SdHO. In contrast to the SdHO contribution to $\sigma^{(2)}_{\alpha 
\alpha}$ [see Eq. (42)], which is proportional to ${\cal J}_{0}$, this term is proportional to ${\cal J}_{2}$.
The contribution with ${\cal J}_{0}$, which describes the average density of states, does not enter the 
SdHO term in Eq. (57) because in view of the assumed isotropy of scattering the average scattering 
rate $a^{-1} \int_0^a dX 2 {\rm Im} \Sigma^{A}_{\varepsilon X}$ depends on $\varepsilon$ exactly in the 
same way as the average density of states, so the corresponding quantum terms in the nominator and 
denominator of Eq. (49) compensate each other, making the first term in Eq. (56) equal to zero at $l=0$.   
As a result, the SdHO in $\sigma^{(1)}_{yy}$ and $\sigma^{(2)}_{\alpha \alpha}$ are shifted in phase by 
$\pi$ in the region ${\tilde u}_Q > \omega_c$, where ${\cal J}_{0}$ and ${\cal J}_{2}$ oscillate in antiphase.
Finally, the term $\delta \sigma^{(1)}_{yy}$ describes the second-order ($\propto d^2$) quantum correction:
\begin{eqnarray}
\delta \sigma^{(1)}_{yy} = \sigma^{(1c)}_{yy} \frac{2d^2 ({\cal S}_1-{\cal S}_2)}{1-J_0^2(QR)}, \\
{\cal S}_1= \sum_{n=1}^{\infty} \left\{ {\cal J}_{n-1} [1-J_0^2((n-1)QR)] \right. \nonumber \\ 
\left. + {\cal J}_{n+1} [1-J_0^2((n+1)QR)] \right\}^2 \frac{J_0^2(nQR)}{1- J_0^2(nQR)}, \nonumber \\
{\cal S}_2=\sum_{n=-\infty}^{\infty} \left\{ {\cal J}^2_{n} J_0^2(nQR)[1-J_0^2(nQR)] \right. \nonumber \\
+ \left. {\cal J}_{n} {\cal J}_{n+2} J_0^2((n+2)QR) [1-J_0^2(nQR)] \right\}. \nonumber
\end{eqnarray} 
The contribution ${\cal S}_1$ is obtained from the last term of $F_1$ [see Eq. (55)] by substituting 
there the first ($\propto d$) term of $\delta \mu$ and then retaining the terms that do not contain 
rapid oscillations with energy, while ${\cal S}_2$ is obtained from the second term of $F_1$ by 
substituting the second ($\propto d^2$) term of $\delta \mu$. Both these contributions are equally 
important. 

As discussed in Sec. II, if the modulation length is small enough, the correlation function 
$C$ can be approximated by the bare correlation function $K$. In the classical transport regime,
this requires the condition $QR \gg 1$ so that $J_0^2(QR) \ll 1$. A formal substitution of 
$K_{\varepsilon}(q,q')$ in place of $C_{\varepsilon}(q,q')$ allows one to skip consideration 
of the integral equation (46). As a result, $\sigma^{(1)}_{yy}$ is presented in a closed 
analytical form:
\begin{eqnarray}
\sigma^{(1)}_{yy} = \frac{e^2 \ell^4 u^2 Q^2}{\pi w} \int d \varepsilon \left(-\frac{\partial f_{\varepsilon}}{\partial \varepsilon} \right) J^2_0(QR_{\varepsilon}) \nonumber \\ 
\times \frac{1}{a} \int_0^a dX \sin^2(QX) \mu(X), 
\end{eqnarray}    
where the harmonic modulation is already assumed.
By employing the density of states $\rho_{\varepsilon X}=(m/\pi)D_{\varepsilon}(X)$, 
the scattering time $\tau_{\varepsilon X}=1/2 {\rm Im} \Sigma^{A}_{\varepsilon X}$, and 
the group velocity $v_{\varepsilon X}= \partial U_{\varepsilon X}/\partial p= 
uJ_0(QR_{\varepsilon})Q \ell^2 \sin(QX)$ (recall that $X=-\ell^2 p$ and that 
$v_{\varepsilon X}$ is equivalently described as the average drift velocity [2]), 
one may rewrite Eq. (59) in a more general way:
\begin{eqnarray}
\sigma^{(1)}_{yy} =  \frac{e^2}{a} \int_0^a dX  
\int d \varepsilon \left(-\frac{\partial f_{\varepsilon}}{\partial \varepsilon} 
\right) \rho_{\varepsilon X} v^2_{\varepsilon X} \tau_{\varepsilon X}. 
\end{eqnarray} 
Therefore, under the approximation $C \simeq K$ the contribution $\sigma^{(1)}$, similar as $\sigma^{(2)}$, 
is presentable as an average of a well-defined local conductivity over the modulation period $a$, which 
is consistent with the observation (see the end of Sec. II) that $\sigma^{(1)}$ can be treated 
locally in this case. The conductivity Eq. (60) has the form and the meaning of one-dimensional 
band conductivity, in contrast to the conductivity $\sigma^{(2)}$, which has hopping nature. 
The presentation of the conductivity as a sum of the band contribution and hopping 
contribution is in use since the earliest theoretical works on modulated 2D electron 
gas [3-5]. The conductivity Eq. (60) is identified with the band contribution obtained in 
the previous theories; for a more direct correspondence one may replace the integral over 
energy by the sum over Landau levels according to $\int d \varepsilon \rho_{\varepsilon}(X) ... 
= (\pi \ell^2)^{-1} \sum_N ...$ with $\varepsilon=\varepsilon_N+U_{NX}$. 
   
Equation (60) can be used for description of the quantum transport regime by taking into account 
rapid oscillations of $\rho_{\varepsilon X}$ and $\tau_{\varepsilon X}$ with energy due to Landau 
quantization. It remains to discuss whether the approximation $C_{\varepsilon}(q,q') \simeq 
K_{\varepsilon}(q,q')$ leading to this equation is justified in the quantum transport regime. 
For description of SdHO, this approximation is applicable at $J_0^2(QR) \ll 1$, similar as in 
the classical regime. However, the second-order quantum contribution $\delta \sigma^{(1)}_{yy}$ 
is beyond the accuracy of this approximation, because $\delta \sigma^{(1)}_{yy}$ contains an 
extra smallness of the order $J_0^2(QR)$, see Eq. (58). Therefore, the consideration of the 
integral equation (46) is indispensable even at $QR \gg 1$.

At low magnetic fields and small modulation periods, the quantum contribution to $\sigma^{(2)}$
dominates both in $\sigma_{xx}$ and $\sigma_{yy}$. However, since the quantum contribution to 
$\sigma^{(1)}_{yy}$ is proportional to the classical conductivity $\sigma^{(1c)}_{yy}$ (see also Ref. [43]), 
it becomes larger than the quantum contribution to $\sigma^{(2)}$ as the magnetic field increases.

\section{Numerical results and discussion}

It is important to plot the magnetic-field dependence of the resistivity in order to demonstrate the 
essential features of the linear response. Figures 4-6 show the resistivity of periodically modulated 
2D electron gas in GaAs quantum wells, where $m$ is equal to 0.067 of the free electron mass. The 
electron density $n_s=5 \times 10^{11}$ cm$^{-2}$ and mobility $3 \times 10^{5}$ cm$^2$/V s are 
chosen, which corresponds to parameters of the experiment of Ref. [10]. The calculations are 
carried out for the case of harmonic modulation, by using Eqs. (42), (57), (58), and (23), the 
latter defines a relation between the resistivity and the conductivity contributions calculated 
in the previous section. The scattering time $\tau$ entering the prefactors $\sigma_d$ in Eq. (42) 
and $\sigma^{(1c)}_{yy}$ in Eqs. (57) and (58) is derived directly from the mobility, while the 
scattering time entering the Dingle factor $d$ is assumed to be 5 times smaller than $\tau$, to 
account for a considerable difference between the transport time and the quantum lifetime typical 
for 2D systems [55]. The results for three different periods $a$ are shown. The modulation is 
assumed to be strong enough so that the amplitude of $U(x)$ considerably exceeds the cyclotron energy 
in the quantum transport region of $B$: $\eta=0.1$ for $a=0.45$ $\mu$m and $\eta=0.15$ for 
$a=1$ $\mu$m and $a=3$ $\mu$m. In each of these figures, two components of the resistivity are 
shown at $T=1$ K and at $T=10$ K. 

\begin{figure}[ht]
\includegraphics[width=9.3cm]{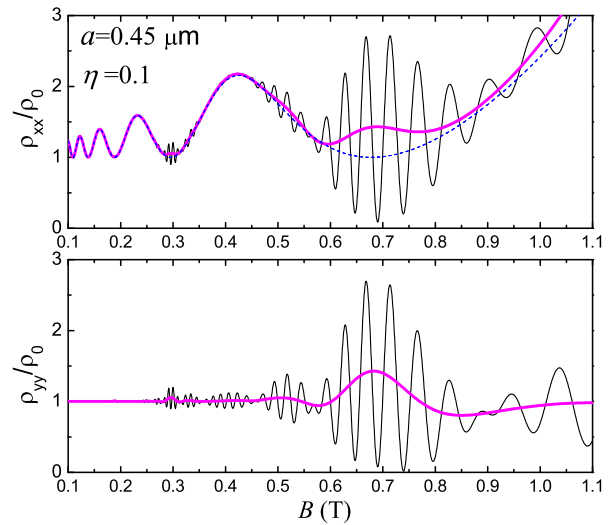}
\caption{(Color online) Resistivity as a function of magnetic field calculated for GaAs quantum well with 
electron density $n_s=5 \times 10^{11}$ cm$^{-2}$ and mobility $3 \times 10^{5}$ cm$^2$/V s for modulation 
strength $\eta =0.1$ and period $a=0.45$ $\mu$m. The upper part shows $\rho_{xx}$ and the lower one 
shows $\rho_{yy}$ expressed in units of zero-field resistivity $\rho_0=m/e^2 n_s \tau$. Thin (black) 
lines correspond to $T=1$ K, thick (colored) lines to $T=10$ K, and the dashed line in the upper part 
shows the classical contribution containing the Weiss oscillations.} 
\end{figure}

\begin{figure}[ht]
\includegraphics[width=9.3cm]{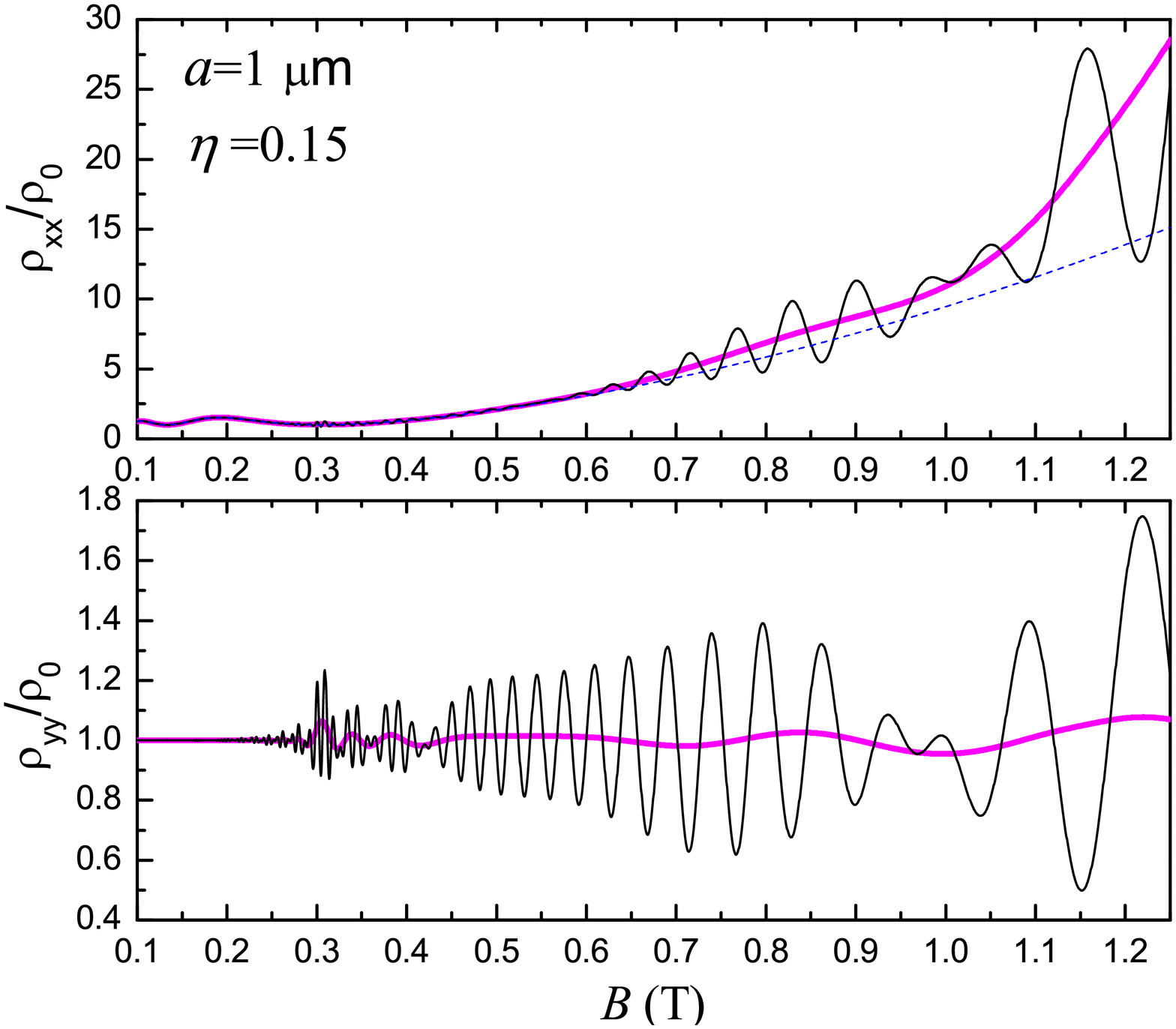}
\caption{(Color online) The same as in Fig. 4 for $\eta =0.15$ and $a=1$ $\mu$m.} 
\end{figure} 

\begin{figure}[ht]
\includegraphics[width=9.3cm]{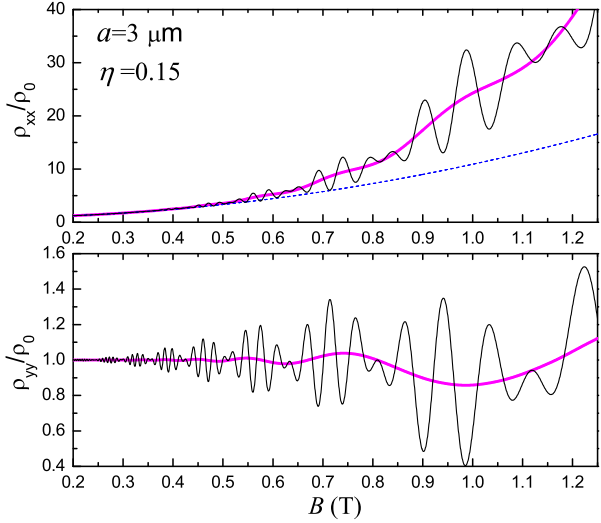}
\caption{(Color online) The same as in Fig. 4 for $\eta =0.15$ and $a=3$ $\mu$m.} 
\end{figure} 

For low temperature, both $\rho_{xx}$ and $\rho_{yy}$ in each of the plots show the SdHO, which are 
significantly modified by the presence of the periodic potential. The basic properties of these 
oscillations have been already explored for the systems with the periods $a \leq 1$ $\mu$m typical 
for experiments on modulated 2D electron gas, and a qualitative agreement between experiment and 
theory is demonstrated [26,27,33,43]. The most important feature is the nonmonotonic dependence of the 
SdHO amplitude on the magnetic field, originating from the modulation of the density of states by 
the periodic potential and formally described by the slowly oscillating factors ${\cal J}_{n}$ defined 
by Eq. (43) and entering the SdHO terms in Eqs. (42) and (57). An indication of such a behavior 
is also seen in an earlier experiment [10] on a system with $a = 1$ $\mu$m. Similar behavior is 
expected for $a = 3$ $\mu$m (Fig. 6). One more interesting feature that follows from the present 
theory is the phase inversion of the SdHO in $\rho_{xx}$. For small-period systems and at low 
$B$, the SdHO of $\rho_{xx}$ and $\rho_{yy}$ are in phase, because they are both determined mostly 
by the contribution $\sigma^{(2)}$. For large-period systems and at higher $B$, the SdHO of 
$\rho_{xx}$ and $\rho_{yy}$ are in antiphase, because $\rho_{xx}$ is now determined by $\sigma^{(1)}_{yy}$, 
while $\rho_{yy}$ is again determined by $\sigma^{(2)}_{xx}$. The origin of the phase shift between 
the SdHO contributions in $\sigma^{(1)}$ and $\sigma^{(2)}$ is explained in the previous section. 

As the temperature increases and the thermal damping factor ${\cal T}_1$ becomes small, the SdHO 
terms are suppressed and the quantum contribution to resistivity is determined by the terms $\delta 
\sigma^{(1)}_{yy}$, $\delta \sigma^{(2)}_{yy}$, and $\delta \sigma^{(2)}_{xx}$. In this case, the 
resistivity, apart from the classical Weiss oscillations, shows long-period oscillations of quantum 
origin. For small-period systems (Fig. 4), the quantum corrections to $\rho_{xx}$ can lead, 
depending on parameters, either to a flattening of the minima of Weiss oscillations or even 
to weak bumps inside these minima, as shown in Fig. 4. Apparently, experimental indications of 
such a behavior are seen in experiments [27,33], though in view of weakness of these features 
one cannot make a definite conclusion on this subject. The effect of enhancement of resistivity 
near the minima of Weiss oscillations takes place because the amplitude of the average potential, 
${\tilde u}_Q$, goes to zero in these minima, and the positive quantum correction  
$\delta \sigma^{(2)}_{yy}$ is no longer suppressed by the modulation. The experimentally observed 
enhancement of SdHO around these minima, discussed in Refs. [27,33], occurs for a similar reason. 
In general, the high-temperature quantum oscillations of $\rho_{xx}$ correlate with the amplitude 
modulation of SdHO as their minima coincide with the nodes of SdHO pattern. This behavior is most 
clearly seen at higher magnetic fields in the system with $a = 3$ $\mu$m (Fig. 6), when the 
quantum correction in $\rho_{xx}$ is determined by the contribution $\delta \sigma^{(1)}_{yy}$. 
Thus, the origin of these oscillations is traced directly to the slow oscillations of the density 
of states caused by the modulation. Formally, according to Eq. (58), the oscillating behavior of 
$\delta \sigma^{(1)}_{yy}$ at $R < a$ is determined by a sum of the terms quadratic in the Bessel 
functions ${\cal J}_{n}$ of the same parity. The terms with even $n$ prevail in this sum, so the 
oscillations basically follow the behavior of ${\cal J}^2_{0}$ and have minima under conditions 
$2 {\tilde u}_Q/\omega_c \simeq l-1/4$ with integer $l$, corresponding to zeros of ${\cal J}_{0}$, 
when the principal oscillating contribution to the density of states is suppressed, see Eq. (38). 
In the limit $QR \ll 1$, when $\tilde{u}_Q$ is independent of $B$ and equal to $u$, the oscillations 
are periodic in $B^{-1}$. The effect of a periodic spatial modulation on the density of states of 2D 
electrons was previously discussed [26,27,33,43] in connection with amplitude modulation of SdHO. The 
present study shows that this effect also leads to long-period resistivity oscillations in $\rho_{xx}$ 
that survive at high temperatures. The existence of such oscillations is not surprising, because the 
modification of the density of states occurs at an energy scale much larger than the cyclotron 
energy, and, therefore, is robust to thermal averaging.
  
The component $\rho_{yy}$ also demonstrates oscillating behavior at high temperatures, though 
it is more complicated and requires a different explanation. In the lower part of Fig. 5, one 
can see two types of oscillations, the short-period ones in the region of low $B$ and the 
long-period ones at higher $B$, which have different origin. The first type of oscillations 
has been observed and theoretically reproduced within a simple model assuming that the 
conductivity is proportional to the integral of the squared average density of states over 
energy [27]. Below, more details are added to explanation of this phenomenon. The low-$B$ 
quantum oscillations, similar to the oscillations of $\rho_{xx}$, appear because of the 
modification of the density of states by the modulation. However, in contrast to the 
oscillations of $\rho_{xx}$, they exist only in the region of low $B$, where $QR \gtrsim 1$. 
The reason for this can be understood by noticing that $\rho_{yy} \propto \sigma^{(2)}_{xx}$ 
is caused by the scattering-assisted hopping transitions, and, therefore, $\rho_{yy}$ is 
proportional to the product of the densities of the states between which the transition takes place, 
see Eq. (39). As the magnetic field varies, $\sigma^{(2)}_{xx}$ oscillates each instant when the maxima 
of the density of states belonging to different Landau subbands are aligned. The density of 
states has maxima at the top and bottom edges of the subbands, this is a general consequence of 
the parabolic dependence of $U_{N X}$ on $X$ near $X=l \pi/Q$, leading to van Hove singularities 
of the density of states in one-dimensional subbands [7,11,33] in the collisionless limit. Thus, the Landau 
subband width $2 \tilde{u}_Q$ plays the role of the resonance energy, and the oscillations of $\rho_{yy}$ 
are periodic in $2 \tilde{u}_Q/\omega_c$, one period corresponds to a change of this ratio by 
unity. However, the resonance hopping between the upper and lower edges 
of two different Landau subbands cannot occur if the maximal hopping distance, equal to the cyclotron 
diameter $2R$, is smaller than the modulation half-period $a/2$. For the case of $a=1$ $\mu$m shown 
in Fig. 5, such a cutoff corresponds to $B \simeq 0.5$ T. Indeed, this is the field when the low-$B$ 
oscillations of $\rho_{yy}$ in Fig. 5 disappear and another type of oscillations emerges. These 
new oscillations are better seen in the case of $a=3$ $\mu$m (the lower part of Fig. 6), 
when the condition $QR < 1$ is realized at $B > 0.25$ T. The oscillations do not correlate 
with the amplitude modulation of the SdHO, so they are not periodic 
in $2 \tilde{u}_Q/\omega_c$. Their period increases with increasing $B$ faster than the period of 
the oscillations of $\rho_{xx}$. To explain the origin of these oscillations, it is again essential 
to recall that the quantum contribution to $\rho_{yy}$ is determined by the hopping transitions 
and that the hopping distance is an important parameter of the transport. 
In the regime of high Landau levels, $N \gg 1$, when electron motion can be treated quasiclassically, 
the probability of such transitions has a maximum when the hopping distance is equal to 
the cyclotron diameter, which corresponds to a backscattering of the electron rotating in a
cyclotron orbit. The property of enhanced backscattering probability in 2D systems is a purely 
kinematic effect, which is not related to the presence of magnetic field. Being combined with the 
cyclotron motion, Landau quantization, and spatial dependence of the potential energy, this property 
leads to oscillating behavior of the resistance, which becomes most clear in the case of $U(x)=|e| E x$, 
when the electric field $E$ is constant. Then, the states with the same energy in different Landau 
levels $N$ and $N'$ are separated by the distance $|N-N'|\omega_c/|e| E$. When this distance is equal 
to $2R$, a resonance takes place, so the conductivity oscillates each instant when the ratio 
$2 |e| E R/\omega_c$ is changed by unity. The resonance energy $2 |e| E R$ is the variation of the 
potential energy on the distance of cyclotron diameter. For arbitrary potential and in the adiabatic limit, when 
$R$ is much smaller than the modulation length, the resonance energy is expressed through the local electric 
field and is equal to $2 |e| E({\bf r}) R$. This resonance effect leads to the oscillating quantum correction 
to the local conductivity given by Eqs. (44) and (45). In periodically modulated systems, such oscillations 
survive after averaging of the local conductivity over the modulation period, though their amplitude 
becomes smaller, and the conductivity oscillates as a function of $2 |e| {\overline E} R/\omega_c$, 
where the parameter ${\overline E}$ approaches to the amplitude of the electric field as the product $QR$ 
decreases. The mechanism discussed above is responsible for high-temperature oscillations of $\rho_{yy}$ 
at large $a$ and $B$. It is also responsible for a special kind of nonlinear phenomena studied in the past 
decades [57-67], in particular, the Hall-field induced resistance oscillations (HIRO), when the 
electric field $E$ that tilts the Landau levels is the Hall field proportional to the electric current. 
Since the Hall field is proportional also to the magnetic field $B$, the product $ER$ is independent 
of $B$, and HIRO are periodic in $B^{-1}$. Thus, the present study demonstrates that the physical 
mechanism that leads to HIRO also produces a special kind of resistance oscillations that can be 
observed in modulated 2D electron systems. In contrast to HIRO, these oscillations exist in the 
linear transport regime and have a well-defined periodicity only in the adiabatic limit $QR \ll 1$,
when they are periodic in $B^{-2}$, because ${\overline E} R \propto B^{-1}$. They should be 
better seen in the systems with a significant amount of short-range scatterers, where backscattering 
is efficient.

In summary, a linear magnetotransport theory of 2D electron gas modulated by a weak and 
classically smooth potential is developed. It is shown that a consistent approach to the 
problem within the quantum linear response (Kubo) formalism requires a consideration of 
nonlocal conductivity tensor. The conductivity tensor is subdivided into the local part, 
$\sigma^{(2)}({\bf r})$, and the nonlocal part, $\sigma^{(1)}({\bf r},{\bf r}')$, proportional 
to a product of the potential gradients in the points ${\bf r}$ and ${\bf r}'$. In the classical 
limit, the local part describes the Drude conductivity, while the nonlocal part is responsible 
for the commensurability oscillations. The nonlocal part in this limit is expressed through the 
correlation functions of a homogeneous 2D electron gas [54]. When Landau quantization becomes 
important, both local and nonlocal parts contain quantum contributions described above for a 
particular case of one-dimensional (unidirectional) periodic modulation. The approximations used 
in the paper include: i) the conditions of quasiclassical transport under classically smooth and 
weak modulation, as summarized in Eq. (2), ii) the condition of classically strong magnetic fields, 
relevant for observation of both commensurability phenomena and quantum oscillations, 
iii) the self-consistent Born approximation, which lefts beyond the effects 
of both weak and strong localization but nevertheless is reasonable for description of transport 
at weak modulation away from the quantum Hall effect regime, and iv) the assumption of isotropic 
scattering, which is not good in application to realistic 2D electron systems with smooth disorder 
[74] but technically necessary in order to obtain a closed equation for the correlation function 
describing the nonlocal conductivity and to express $\sigma^{(1)}$ in the analytical form in the 
classical limit. Even within these approximations, the problem of quantum magnetotransport remains 
complicated, because determination of the correlation functions describing the nonlocal part of the 
conductivity requires a solution of the integral equation, Eq. (46). Evaluation of the local part is 
simpler and leads to the expression Eq. (39) generalizing Titeica's formula for hopping conductivity 
in magnetic field. Analytical expressions for both parts of the conductivity tensor are obtained 
in the case of moderately strong magnetic fields, when the quantum contributions are not 
large and an expansion of the conductivity in powers of the Dingle factors is possible.  
It is found that the Shubnikov-de Haas oscillations coming from the contributions $\sigma^{(1)}$ 
and $\sigma^{(2)}$ have opposite phases. The theory suggests that, apart from the Shubnikov-de Haas 
oscillations modified by the modulation potential, there exist other kinds of quantum oscillations 
with smaller amplitudes, which survive when the temperature increases. The resistance in the direction 
perpendicular to the modulation axis, $\rho_{yy}$, shows two different kinds of such oscillations. For 
detection of this behavior, experimental studies of 2D electron systems with enhanced modulation strength 
(10-15\%) and larger modulation periods (several microns) are desirable, as demonstrated above by the numerical 
calculations. These conditions are required to make the amplitude of the modulation potential $U(x)$ 
considerably larger than the cyclotron energy in the region of $B$ where Landau quantization is 
important, essentially including the region $R < a$, where the quantum oscillations caused by 
enhanced backscattering are expected in $\rho_{yy}$ and the high-temperature quantum oscillations of 
$\rho_{xx}$ are no longer obscured by the presence of Weiss oscillations. In the majority of experiments, 
mostly focused on the magnetotransport in the regime of Weiss oscillations, GaAs samples with the 
period $a$ smaller than 0.5 $\mu$m were used. Only a few experiments [10,27,33] on the samples with 
$a=1$ $\mu$m are available, and the author is not aware of experiments on samples with larger periods. 
While the experiments in Refs. [10,27,33] employ sufficiently strong modulation, the quantum contribution 
to resistance in Ref. [10] is weak, apparently because of the effect of disorder, and the resistance 
in Refs. [27,33], measured for samples with higher mobilities, is shown only in the region of fields 
below 0.5 T, where $R > a$. Thus, the available experimental data does not allow to verify the existence 
of the oscillations discussed in this paper and demonstrated in the high-temperature magnetoresistance 
plots in Fig. 6 and high-field part of Fig. 5. Further experiments are required for this purpose.

\end{document}